\documentclass[]{raa}            % referee version: for submission

\usepackage{graphicx,times}             %for PS/EPS graphics inclusion, new
\usepackage{natbib}
\usepackage{ulem}
\usepackage{cancel}
\usepackage{amssymb,amsmath}
\bibpunct{(}{)}{;}{a}{}{,}

\usepackage[pagebackref=true]{hyperref}
\usepackage{CJK}
\usepackage{soul}
\usepackage{longtable}   % 跨页长表
\usepackage{booktabs}    % 好看横线
\usepackage{array}       % 表格辅助
\begin{document}
\begin{CJK*}{UTF8}{gbsn}
  \title{Newly discovered Luminous blue variable candidates in M31 \& M33}
%   \subtitle{I. Place Your Subtitle Here}

   \volnopage{Vol.0 (20xx) No.0, 000--000}      %%preserved for Editor. DOn't remove!
   \setcounter{page}{1}          %%starting page, preserved for Editor. DOn't remove!

   \author{Sai Li  % Put your Chinese name in "( )" if you like. Note to open line 11 "\usepackage[UTF8]{ctex}"
      \inst{1}
   \and Cheng Liu 
      \inst{1}
   \and Jincheng Guo 
      \inst{1}
   }

   \institute{Department of Scientific Research, Beijing Planetarium, Beijing 100044, People's Republic of China; {\it liucheng@bjp.org.cn (CLiu)}, {\it andrewbooksatnaoc@gmail.com (JCGuo)}\\
\vs\no
   {\small Received 20xx month day; accepted 20xx month day}}

\abstract{This study presents an investigation of nearly two dozen candidate Luminous Blue Variables (cLBVs) in the galaxies M31 and M33. Eight stars have been studied in detail, while an additional sixteen objects are briefly mentioned. Multi-epoch spectra of confirmed cLBVs from LAMOST and previous literature show broad hydrogen, He I lines, abundant Fe II and [Fe II] emission lines, and discernible spectral variability, consistent with the characteristics of known LBVs. Low outflow velocities inferred from P Cygni profiles are also incorporated into the classification criteria. Moreover, key stellar properties, including temperature and luminosity, are determined using the Spectral Energy Distribution (SED) fitting and spectral modeling. By comparison with stellar evolutionary tracks on the temperature–luminosity diagram, the initial masses are estimated to be in the range of approximately 32 to 60\,$\mathrm{\textit{M}_{\odot}}$. Except for J013401 and J013411, other stars locate within the typical LBV region between the S Doradus instability strip and their outburst phase. More importantly, our sample, except for the binary system, are all positioned in the LBVs region rather than that of B[e]SGs in the near-infrared color-color diagram. 
Based on all available information, one of the eight sources is confirmed as an LBV, four stars are designated as high-probability cLBVs, and the remaining three stars await further photometric observations to secure their classification.
Given the current scarcity of known cLBVs, our study has the potential to make a significant increase in the number of LBVs in M31 and M33.
\keywords{galaxies: spiral --- stars: variables: S Doradus --- stars: massive --- stars: winds, outflows}
}

   \authorrunning{S. Li, C. Liu \& J. C. Guo }            %author_head in even pages
   \titlerunning{Newly discovered Luminous blue variable candidates in M31 \& M33 }  % title_head in odd pages
   \maketitle

\section{Introduction}           %% first-level sections will be auto-capitalized
\label{sect:intro}

LBVs are widely recognized as a peculiar type of massive (\textit{M}\,$\geq$ 25\,$\mathrm{\textit{M}_\odot}$; \citealp{2016ApJ...825...64H}) post-main-sequence stars with the main characteristics of high luminosity and obvious spectroscopic and photometric variability ranging from weeks to years \citep{1994PASP..106.1025H,2001A&A...366..508V}. They are also named as S Doradus variables, with 1-2\,mag photometric variation over a period of years to decades, accompanied by mass-loss rates of $\mathrm{\sim}$$\mathrm{ 10^{-5} }$\,-\,$\mathrm{ 10^{-4} }$\,$\mathrm{\textit{M}_\odot} \cdot$yr$\mathrm{^{-1}}$ \citep{2020MNRAS.493.2410W,2018ApJ...864...31A,2019ApJ...884L...7H}. According to \citet{1994PASP..106.1025H}, brightness difference of $\leq$ 1.0 mag may also occur. However, the physical mechanism resulting in LBV variability is still not fully understood. Recently, a model developed by \citet{2021A&A...647A..99G} reproduces the typical observational phenomenology of the S Doradus variability. The S Doradus variability plays extremely important roles in distinguishing LBVs.

In the Hertzsprung-Russell (H-R) diagram, LBVs are located between the main sequence (MS) and the Humphreys-Davidson (H-D) luminosity limit above which only highly unstable objects are found \citep{2020MNRAS.493.2410W}. In line with conventional stellar evolution, LBVs are assumed to be in a brief transitional period between the main sequence O-type stars and the Wolf-Rayet (WR) stars \citep{1994PASP..106.1025H,2000ARA&A..38..143M,2021A&A...654A.167U,2019ApJ...884L...7H,2023MNRAS.518.4345S}. However, over the past decade, the theoretical comprehension of the evolution of massive stars has undergone a radical change \citep{2017RSPTA.37560268S}. 

\begin{figure*}[htbp]
\includegraphics[angle=0,scale=0.25]{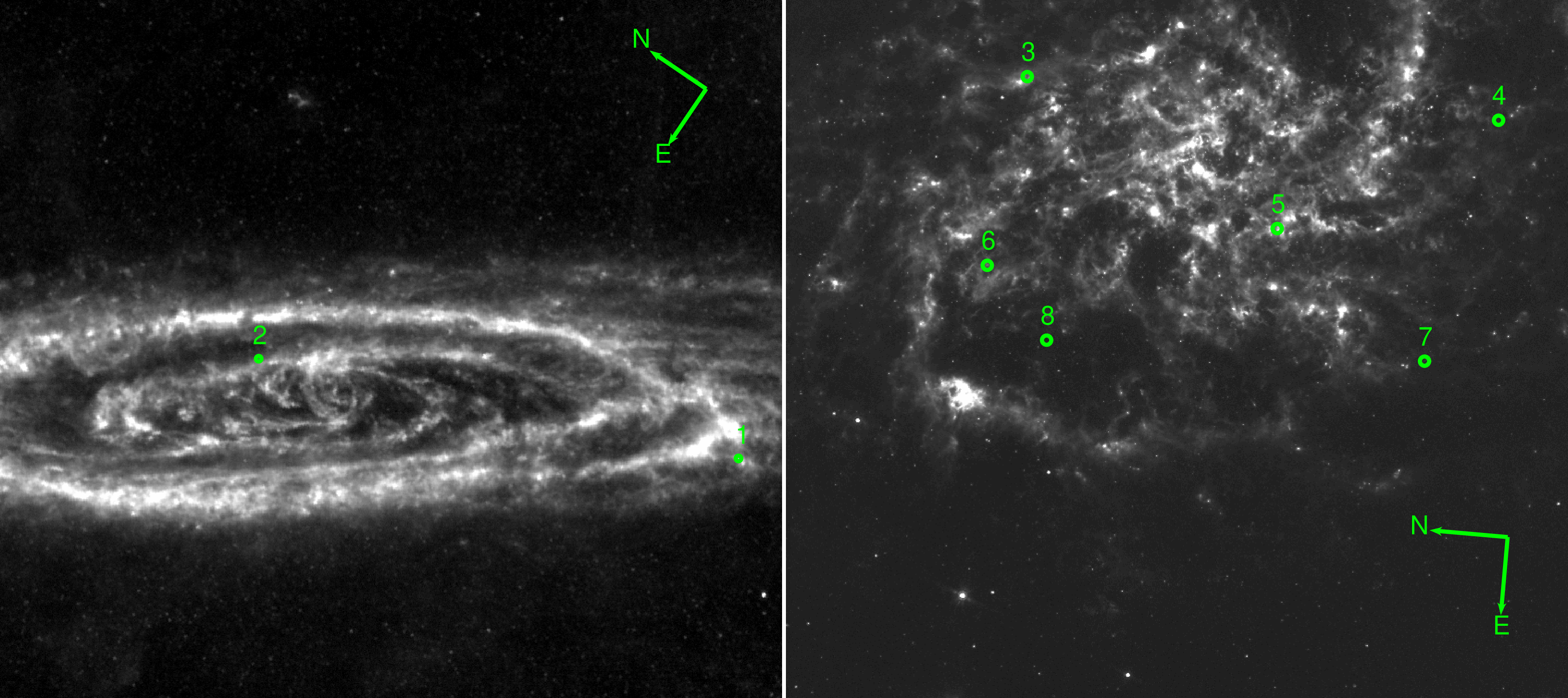}
\centering

\caption{ Herschel SPIRE 250 $\mu$m images. Left: The two green circles mark the locations of our cLBVs in M31.  Right: The six green circles mark the locations of our cLBVs in M33.
}

\label{fig:her}
\end{figure*}

In order to study the physical nature of LBVs, significant research efforts have been dedicated to the discovery of new objects. In recent years, due to many works conducted to study M31 and M33 (e.g.\citealp{2013ApJ...773...46H,2014ApJ...790...48H,2017ApJ...844...40H,2017ApJ...836...64H,2016ApJ...825...50G,2020MNRAS.497..687S,2022RAA....22a5022S}), many LBVs have been confirmed and more cLBVs are investigated. \citet{2015MNRAS.447.2459S} studied five cLBVs in M31. Among them, J004526.62+415006.3 and J004051.59+403303.0 were confirmed as LBVs. The other two candidates, J004417.10+411928.0 and J004444.52+412804.0, were identified as B[e] supergiants (B[e]SGs). The status of J004350.50+414611.4 remains uncertain, as it is still being considered as a LBV candidate. According to \citet{2015PASP..127..347H}, J004526.62+415006.3 in M31 was studied as an LBV. In more previous studies, \citet{2018RNAAS...2..121R} confirmed six LBVs in M31, whose names are LGGS J004051.59+403303.0, AE And, AF And, Var 15, Var A-1, and UCAC4 660-003111; and five LBVs in M33, whose names are Var C, Var B, Var 83, Var 2, and Romano's Star. Furthermore, IR spectra for eight cLBVs in M31 and M33 were published by \citet{2014ApJ...780L..10K} and \citet{2018MNRAS.480.3706K}. \citet{2019ApJ...884L...7H} reported the discovery of a new LBV: LAMOST J0037+4016 in the distant outskirts of M31. More recently, \citet{2020MNRAS.497..687S,2022RAA....22a5022S} studied a LBV, J004341.84+411112.0 in M31 and \citet{2023RNAAS...7...96M} also found a cLBV, M33C-4119 (LGGS J013312.81+303012.6) in M33. As the number of LBVs and cLBVs in M31 and M33 continues to grow, more in-depth research into the physical nature of LBVs becomes feasible. 

In the work of \citet{2022ApJ...932...29L}, there are some sources exhibiting strong emission features and spectral variability over a range of years, which draws our attention. In this work, in order to confirm whether they are rare type star - LBVs and reveal their physical properties, we mainly focus on the spectroscopic and photometric analysis of eight sources in M31 and M33, with two in M31 and six in M33. Among them, a binary system and a confirmed LBV are included. The specific locations of these eight sources within their host galaxy are revealed by the Herschel SPIRE 250\,$\mu$m \citep{2012A&A...546A..34F} images in Figure \ref{fig:her}. Infrared images are selected to better display the structures and locations of sprial arms and molecular cloud of M31 \& M33. Our candidates are located in the spiral arms or overlap with molecular cloud. The selection of our cLBVs sample is motivated by the availability of multi-epoch spectra, which facilitates the comparison of potential spectral variations. And additional sixteen cLBVs with lower probability will be briefly discussed, focusing on their color-color diagram and single spectrum. The photometric and spectroscopic datasets used in this work are described in Section \ref{sec:Data}. The main work and individual physical properties are presented in Section \ref{sec:Result}. We discussed and summarized in Section \ref{sec:conclusion}.

\section{Data}\label{sec:Data} 
To further reveal the physical nature of LBVs, we initially introduce the existing photometric and spectroscopic information, covering both previous and recently acquired data for this study.

\subsection{Photometric Data}
The photometric data of the eight high-probability candidates are primarily collected from the Catalina Real-Time Transient Survey (CRTS; \citealp{2009ApJ...696..870D}) and the All-Sky Automated Survey for Supernovae (ASAS-SN; \citealp{2019MNRAS.486.1907J}). CRTS provided \textit{V}-band observations spanning approximately 8 yrs, from 2005 to 2013, using the 0.7-meter Catalina Schmidt Survey (CSS) telescope situated north of Tucson, Arizona. ASAS-SN contributed a collection of \textit{V}-band measurements across various epochs over a period of roughly 7 yrs, from 2012 to 2018, and mostly $g$-band measurements from 2018 to the present. We applied a 3$\sigma$ clipping to the light curves and excluded photometric measurements with limiting magnitudes greater than 18 mag. Furthermore, this research incorporates photometry from the Zwicky Transient Facility (ZTF; \textit{gri}; \citealp{2020ApJS..251....7F}), covering $\sim$\,7 yrs from 2018 to 2024. In addition to the data from CRTS, ASAS-SN, and ZTF, we also have integrated observations from several other multiband optical/near-infrared photometry surveys. These include the Local Group Galaxy Survey (LGGS; \textit{UBVRI}; \citealp{2006AJ....131.2478M}), Panoramic Survey Telescope and Rapid Response System (Pan-STARRS; \textit{grizy}; \citealp{2018AAS...23110201C}), and Two Micron All-Sky Survey (2MASS; \textit{JH}$\mathrm{\textit{K}_{s}}$; \citealp{2006AJ....131.1163S}). The corresponding magnitudes of all datasets have been calibrated to the Johnson-Cousins photometric systems using a series of transformation formulas (methods are described in the following section). Table \ref{tab:photometry} presents all the photometric observations described above.

\begin{table*}
\centering
\caption{Photometric Information of the eight cLBVs (Sample A).}
\label{tab:photometry}
\begin{tabular}{@{}lcccc@{}}
\noalign{\smallskip}
\noalign{\smallskip}
\hline
\noalign{\smallskip}
Date & Source & Filters & Epochs\\
% \noalign{\smallskip}
\hline
\noalign{\smallskip}
%-------------  左半边  -------------
\multicolumn{4}{c}{\textbf{J004051.59+403303.0 (J004051)}}\\
2000-10-25 & 2MASS    & \textit{JH}\ensuremath{K_s} & 1\\
2000       & LGGS     & \textit{UBVR}             & 1\\
2005-2013  & CRTS-CSS & \textit{V}                & 210\\
2014-06-30 & Pan-STARRS & \textit{grizy}          & 1\\
2013-2024   & ASAS-SN  & \textit{V}\textit{g}    & 64\\
2018-2024  & ZTF      & \textit{gri}              & 746\\
\noalign{\smallskip}
\hline
\noalign{\smallskip}
\multicolumn{4}{c}{\textbf{J004253.41+412700.5 (J004253)}}\\
2000       & LGGS     & \textit{UBVR}             & 1\\
2014-09-20 & Pan-STARRS & \textit{grizy}          & 1\\
2018-2024  & ZTF      & \textit{gri}              & 675\\

\noalign{\smallskip}
\hline
\noalign{\smallskip}
\multicolumn{4}{c}{\textbf{J013339.48+304540.5 (J013339)}}\\
1997-12-05 & 2MASS    & \textit{JH}\ensuremath{K_s} & 1\\
2000       & LGGS     & \textit{UBVR}             & 1\\
2005-2013  & CRTS-CSS & \textit{V}                & 317\\
2014-08-28 & Pan-STARRS & \textit{grizy}          & 1\\
2018-2024  & ZTF      & \textit{gri}              & 254\\
\noalign{\smallskip}
\hline
\noalign{\smallskip}
\multicolumn{4}{c}{\textbf{J013340.05+302845.7 (J013340)}}\\
1997-12-05 & 2MASS    & \textit{JH}\ensuremath{K_s} & 1\\
2000       & LGGS     & \textit{UBVR}             & 1\\
2005-2013  & CRTS-CSS & \textit{V}                & 318\\
2014-07-26 & Pan-STARRS & \textit{grizy}          & 1\\
2013-2024  & ASAS-SN  & \textit{V}\textit{g}                  & 621\\
2018-2019  & ZTF      & \textit{gri}              & 28\\
\noalign{\smallskip}
\hline
\end{tabular}%
\hfill   % 让左右两半之间留空
% \hspace{-10em}%   ← 关键：负间距把两列拉近
\begin{tabular}{@{}lcccc@{}}
% \noalign{\smallskip}
% \noalign{\smallskip} 
\hline
\noalign{\smallskip} 
Date & Source & Filters & Epochs\\
\hline
\noalign{\smallskip}
%-------------  右半边  -------------
\multicolumn{4}{c}{\textbf{J013401.16+303618.4 (J013401)}}\\
1997-12-05 & 2MASS    & \textit{JH}\ensuremath{K_s} & 1\\
2000       & LGGS     & \textit{UBVR}             & 1\\
% \del{2005-2013  & CRTS-CSS & \textit{V}                & 270\\}
2011-2015  & Pan-STARRS & \textit{grizy}          & 3\\
% 2013-2018  & ASAS-SN  & \textit{V}                & 396\\
2018-2019  & ZTF      & \textit{gri}              & 28\\
\noalign{\smallskip}
\hline
\noalign{\smallskip}
\multicolumn{4}{c}{\textbf{J013411.32+304631.4 (J013411)}}\\
1997-12-05 & 2MASS    & \textit{JH}\ensuremath{K_s} & 1\\
2000       & LGGS     & \textit{UBVR}             & 1\\
% \del{2005-2013  & CRTS-CSS & \textit{V}                & 276\\}322.
2014-07-06 & Pan-STARRS & \textit{grizy}          & 1\\
2018-2024  & ZTF      & \textit{gri}              & 244\\
\noalign{\smallskip}
\hline
\noalign{\smallskip}
\multicolumn{4}{c}{\textbf{J013420.91+303039.6 (J013420)}}\\
1997-12-05 & 2MASS    & \textit{JH}\ensuremath{K_s} & 1\\
2000       & LGGS     & \textit{UBVR}             & 1\\
2005-2013  & CRTS-CSS & \textit{V}                & 314\\
2014-08-18 & Pan-STARRS & \textit{grizy}          & 1\\
% 2013-2018  & ASAS-SN  & \textit{V}                & 396\\
2018-2024  & ZTF      & \textit{gri}              & 239\\
\noalign{\smallskip}
\hline
\noalign{\smallskip}
\multicolumn{4}{c}{\textbf{J013422.87+304410.8 (J013422)}}\\
2000       & LGGS     & \textit{UBVR}             & 1\\
2005-2013  & CRTS-CSS & \textit{V}                & 206\\
2011-2014  & Pan-STARRS & \textit{grizy}          & 2\\
2013-2024  & ASAS-SN  & \textit{V}\textit{g}                   & 615\\
2018-2024  & ZTF      & \textit{gri}              & 254\\
% \noalign{\smallskip}
% \hline
\bottomrule
\end{tabular}
\end{table*}

\subsection{Spectroscopic Data}
The majority of our sample spectra are acquired from LAMOST Spectroscopic Survey \citep{2012RAA....12..735D,2014IAUS..298..310L}. LAMOST can simultaneously collect up to 4000 optical ($\lambda$\,$\sim$\,3700\,\AA\,-\,9000\,\AA), low-resolution (R\,$\sim$\,1800) spectra in one exposure. The data were processed using the LAMOST standard pipeline \citep{2012RAA....12.1243L}. Moreover, we analyzed comparative spectral data from \cite{2012ApJ...759...11N} and \cite{2017ApJ...836...64H}, acquired in October 2010 with the Hectospec multi-object spectrograph \citep{1998SPIE.3355..285F,2005PASP..117.1411F} on the 6.5-m MMT telescope at Mount Hopkins. The Hectospec possesses a 1-degree field of view and is equipped with 300 optical fibers. The research in \cite{2013ApJ...773...46H} utilized a 600 line per millimeter grating with a 4800\,\AA\,tilt, providing a spectral range from 3550\,\AA\,to 6050\,\AA\,with a resolution of 0.54\,\AA\,per pixel or R\,$\sim$\,2000. For the red spectral region, the same grating was employed with a 6800\,\AA\,tilt, achieving a spectral range from 5550\,\AA\,to 8050\,\AA, resolution of 0.54\,\AA\,per pixel or R\,$\sim$\,3600. The blue and red spectral data were processed using an exportable version of the CfA/SAO SPECROAD software package designed for Hectospec data reduction. A comprehensive overview of the spectroscopic observations is presented in Table \ref{tab:spec}.

% 自定义表头
\newcommand{\mytablehead}{%
  \hline
  Date & Seeing & Total exp. & Spectral Resolution (FWHM) & Spectral range & Source\\
       & (arcsec) & (s) & (\,\AA) & (\,\AA) & \\ 
  \hline
}

%---------------- 正文 ----------------
\small
\setlength{\tabcolsep}{4pt}
\renewcommand{\arraystretch}{1.1}

\begin{longtable}{@{}cccccc@{}}
\caption{Spectroscopic Information of the eight cLBVs (Sample A).}\\

\mytablehead
\endfirsthead
\label{tab:spec}

\endfoot

\hline
\endlastfoot

%========= 全部内容 =========
\noalign{\smallskip}
\multicolumn{6}{c}{\textbf{J004051}}\\
2011-10-21 &--&3600&4.5--5.2&3700--9000&Hectospec\\
2011-10-24 & 3 & 3600 & 2.5 & 3700--9000 & LAMOST\\
2011-11-17 &--&5400&4.5--5.2&3700--9000&Hectospec\\
2013-10-09 &--&5400&2.0--2.2&5550-8050&Hectospec\\
2013-10-12 &--&7200&2.0--2.2&3550--6050&Hectospec\\
2015-09-20 & -- & 3600 & 2.1 & 3550--6050 & (1)\\

\hline
\noalign{\smallskip}
\multicolumn{6}{c}{\textbf{J004253}}\\
2011-10-21 &--&3600&4.5--5.2&3700--9000&Hectospec\\
2011-10-28 & 2.8 & 2700 & 2.5 & 3700--9000 & LAMOST\\
% 2013-09-25 &&&&&Hectospec\\
2013-09-25 & -- & 9000 & 2.1 & 3550--6050 & (1)\\
2013-09-26 & -- & 5400 & 2.0--2.2 & 5550--8050 & (1)\\
2013-10-16 & 3.5 & 1800 & 2.5 & 3700--9000 & LAMOST\\

\hline
\noalign{\smallskip}
\multicolumn{6}{c}{\textbf{J013339}}\\
2010-10-03 & -- & 7200 & 2.1 & 3550--6050 & (1)\\
2010-10-03 & -- & 5255 & 2.2--2.0 & 5550--8050 & (1)\\
2010-11-26 &--&8100&2.1&3550--6050&Hectospec\\
2012-11-06 &--&5400&4.5--5.2&3700--9000&Hectospec\\
2012-11-07 &--&5400&4.5--5.2&3700--9000&Hectospec\\
2012-11-08 &--&1953&4.5--5.2&3700--9000&Hectospec\\
2012-12-08 &--&4500&4.5--5.2&3700--9000&Hectospec\\
2012-12-08 &--&4500&4.5--5.2&3700--9000&Hectospec\\
2013-10-07 & -- & 9000 & 2.1 & 3550--6050 & (1)\\
2013-10-07 & -- & 7200 & 2.0--2.2 & 5550--8050 & (1)\\
2018-10-09 & 3.4 & 5400 & 2.5 & 3700--9000 & LAMOST\\

\hline
\noalign{\smallskip}
\multicolumn{6}{c}{\textbf{J013340}}\\
2011-12-11 & 3.1 & 1800 & 2.5 & 3700--9000 & LAMOST\\
2014-12-01 & 4.6 & 1500 & 2.5 & 3700--9000 & LAMOST\\

\hline
\noalign{\smallskip}
\multicolumn{6}{c}{\textbf{J013401}}\\
2010-10-03 & -- & 7200 & 2.1 & 3550--6050 & (1)\\
2010-10-03 & -- & 5255 & 2.2--2.0 & 5550--8050 & (1)\\
2013-10-07 & -- & 9000 & 2.1 & 3550--6050 & (1)\\
2013-10-07 & -- & 7200 & 2.2--2.0 & 5550--8050 & (1)\\
2018-10-09 & 3.4 & 5400 & 2.5 & 3700--9000 & LAMOST\\

\hline
\noalign{\smallskip}
\multicolumn{6}{c}{\textbf{J013411}}\\
2009-12-15 &--&3600&2.1&6550-9050&Hectospec\\
2011-12-07 & 4.8 & 900 & 2.5 & 3700--9000 & LAMOST\\
2011-12-11 & 3.1 & 1800 & 2.5 & 3700--9000 & LAMOST\\

\hline
\noalign{\smallskip}
\multicolumn{6}{c}{\textbf{J013420}}\\
2009-12-27 & -- & 3600 & 2.1 & 6550--9050 & (1)\\
2012-01-04 & 3.4 & 2700 & 2.5 & 3700--9000 & LAMOST\\
2012-01-13 & 2.5 & 1800 & 2.5 & 3700--9000 & LAMOST\\
2014-11-16 & -- & 7200 & 2.1 & 3550--6050 & (1)\\
2014-11-29 & -- & 5400 & 2.2--2.0 & 5550--8050 & (1)\\

\hline
\noalign{\smallskip}
\multicolumn{6}{c}{\textbf{J013422}}\\
2010-10-03 & -- & 7200 & 2.1 & 3550--6050 & (1)\\
2010-10-03 & -- & 5255 & 2.2--2.0 & 5550--8050 & (1)\\
2010-10-03 &--&5255&2.0--2.2&5550--8050&Hectospec\\
2010-10-09 &--&8100&2.1&3550--6050&Hectospec\\
2011-12-25 & 2.9 & 2400 & 2.5 & 3700--9000 & LAMOST\\
\hline

% \begin{tablenotes}

% \end{tablenotes}
\end{longtable}
Source: (1) \citet{2017ApJ...836...64H}.

\section{Methodology} \label{sec:Result}
\subsection{Photometry}
Firstly, measurements in other systems have been converted to Johnson \textit{V} band magnitudes to obtain the light curves of these eight objects. In terms of Pan-STARRS and ZTF filter systems, the method mentioned in \citet{2018BlgAJ..28....3K} is used to calculate Johnson \textit{V} magnitudes, 
\begin{equation}
   \mathrm{\textit{V}=\textit{g}-0.508\times(\textit{g}-\textit{r})-0.017.} 
\label{con:Mni}
\end{equation}
For the CRTS-CSS, Johnson \textit{V} band magnitudes are transformed using the equation of \citet{2015Natur.518...74G},
\begin{equation}
    \mathrm{\textit{V}=\mathrm{\textit{V}_{CSS}}+0.31\times(\textit{B}-\textit{V})^2+0.04,} 
\label{con:Mni1}  
\end{equation}
The average \textit{B}\,-\,\textit{V} color is calculated using photometry from LGGS, Pan-STARRS, and ZTF. For the latter two surveys, the \textit{B}\,-\,\textit{V} values are obtained by converting their \textit{g}\,-\,\textit{r} colors to the Johnson-Cousins system using the transformation relations provided by \citet{2018BlgAJ..28....3K}. The \textit{V} band magnitudes provided by LGGS and ASAS-SN are already in the Johnson system and the \textit{g} band magnitudes provided by ASAS-SN have been transformed into the Johnson system using Formula \ref{con:Mni}. After the correction, the data from CRTS-CSS with deviations exceeding 3\,$\sigma$ have been excluded, see Figures in Appendix \ref{sec:phot}.

\subsection{Spectroscopy}
\begin{figure}[htbp]
\includegraphics[angle=0,scale=0.7]{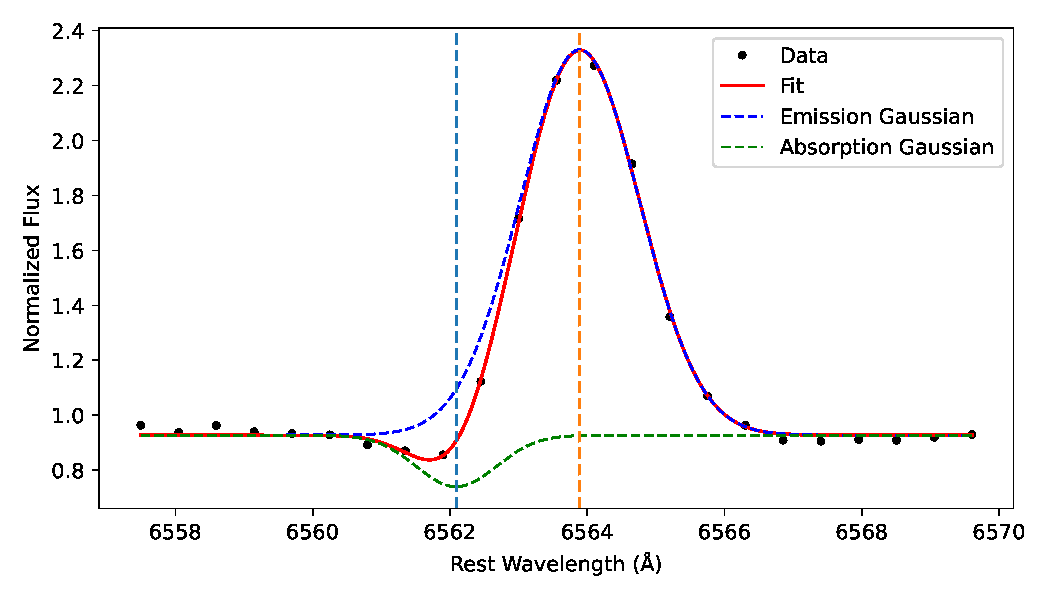}
\centering
\caption{An example of H$\alpha$ P Cygni profile double Gaussian fitting for J013420. Black points show the observed spectrum; the red solid line is the overall Gaussian fit. The blue and green dashed lines represent Gaussians correspond to the emission and absorption components, respectively. The shallow blue and orange vertical dashed lines mark the absorption and emission minimum, respectively.
}

\label{fig:pcygni}
\end{figure}

 In addition to the light curves, we have collected available spectra of these eight candidates from LAMOST, Hectospec archive and \citet{2017ApJ...836...64H}. These spectra are shown in Appendix \ref{sec:spectra}. Series of primary lines are identified and labeled, including Balmer emission lines, He I, Fe II, [Fe II], [Si II], [N II] and so on. The typical spectral lines of well-studied LBV stars include broad and bright emission lines of hydrogen and He I with P Cygni profiles, emission lines of Fe II, as well as forbidden lines of [Fe II] and [N II] \citep{2020MNRAS.497.4834S}. Lines of [O I] $\lambda$$\lambda$ 6300, 6364\,\AA \,are absent among confirmed LBVs. More interestingly, as \citet{2017ApJ...836...64H} mentioned that confirmed LBVs have relatively low stellar wind speeds, normally\,$\sim$\,100\,-\,200\,km $\mathrm{s^{-1}}$, in their hot, quiescent or visual minimum state, compared to the B-type supergiants and Of/WN stars. Moreover, multi-epoch spectra are adopted to identify the spectral variability ranging from weeks to years. %Consequently, to further classify whether our studied sources are LBVs, spectral analysis (about existence of lines and spectral changes) and measuring wind speeds are crucial in the following work. 
 Stars exhibiting P Cygni profiles in their hydrogen and He I lines are listed in Table~\ref{tab:vterm}, together with their outflow velocities derived from the P Cygni profile. Figure~\ref{fig:pcygni} presents an example (J013420) of the double Gaussian-fitted P Cygni profile used for the measurement. More details will be discussed individually in Section \ref{sec:individual}.
% \section*{B}

\subsection{The spectral energy distribution}\label{sed}

To infer additional physical properties of these cLBVs, we constructed their SEDs across multi-band photometric data from the ultraviolet (UV) to the infrared (IR). It is important to note that the physical parameters derived from this method are often subject to inherent degeneracy between reddening and temperature, a well-documented phenomenon in astrophysics. However, this degeneracy can be mitigated by considering the unique evolutionary behavior of LBVs: as these stars evolve, they maintain a nearly constant bolometric luminosity while their effective temperatures change, either becoming cooler and brighter or hotter and fainter \citep{2015MNRAS.447.2459S}.

Using the model BT-Settl-AGSS2009 \citep{2013MSAIS..24..128A} within the SED analysis tool VOSA v7.0 \citep{2008A&A...492..277B}, the SEDs of these eight candidates were fitted. For sources exhibiting significant IR excess, an additional blackbody component was introduced to reproduce the excess emission.  However, due to contamination from a nearby source in the vicinity of J013401, its SED analysis is unreliable. Therefore, it is excluded from the fitting analysis. The SED plots for the remaining objects are presented in Appendix \ref{sec:specfit}, and the derived parameter values for them are summarized in Table \ref{tab:color}.

\begin{table}
\caption{Outflow velocities (km s$^{-1}$).}
\centering
\label{tab:vterm}

\begin{tabular}{cccccc} 
\noalign{\smallskip}
\noalign{\smallskip}
         \hline
  
\noalign{\smallskip}
% \hline

 Name&P Cygni (H $\alpha$)&P Cygni (H $\beta$)&P Cygni (H $\gamma$)&P Cygni (H $\delta$)&P Cygni (He I)\\
         % \noalign{\smallskip}
         \hline
            \noalign{\smallskip}
              J004051& --252.42&--183.07&--153.68&...&...\\
J004253&...&--119.02&--146.61&--134.19&--51.21\\
J013339&...&--148.80&...&--108.73&--94.33\\
      % J013340&...&...&...&...&...\\
       J013401&&&--158.82&&--67.56\\
      J013411&242.98&...&...&...&...\\
      J013420&--115.00&...&...&...&...\\
      J013422&--107.85&...&...&...&...\\
 \noalign{\smallskip}
         \hline
            \noalign{\smallskip}
    \end{tabular}

\end{table}

\subsection{Spectral modeling}\label{spectra_fit}
\label{specm}
In this work, a tool called \textit{RVSpecFit} \citep{2011ApJ...736..146K} is employed to determine stellar atmospheric parameters from spectra through direct pixel fitting using interpolated stellar templates. The tool primarily utilizes the \textit{PHOENIX} \citep{2013A&A...553A...6H} libraries of synthetic stellar atmospheres. To achieve more accurate estimation of the stellar parameters, we modeled multi-spectra covering wavelength from 3700\,\AA\, to \,6500\,\AA. During the modeling process, the H$\beta$ emission line was intentionally excluded to avoid its potential influence on the results. It should be noticed that not all spectra were fitted and gave reliable stellar parameters because of strong emission lines or (and) low SNR. The final best-fitting model of LAMOST spectra is presented in Appendix \ref{sec:specfit}, and the derived parameters are listed in Table \ref{tab:color}. Clearly, the temperature obtained from this procedure is consistent with that obtained from SED within uncertainties.

\subsection{Near-infrared Color-Color Diagram}

\begin{figure}[htbp]
\includegraphics[angle=0,scale=1.2]{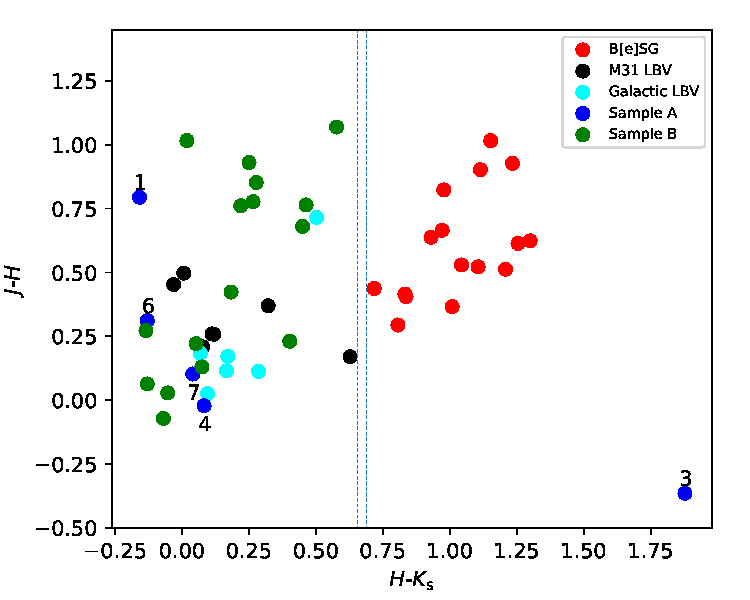}
\centering
\caption{\textit{H}\,-\,$\mathrm{\textit{K}_s}$ vs. \textit{J}\,-\,\textit{H} color-color diagram of LBVs and B[e]SGs. Galactic LBVs and B[e]SGs are shown by cyan and red filled circles, respectively. Confirmed M31 LBVs are shown by black filled circles. The datasets of B[e]SGs, Galatic LBVs and confirmed M31 LBVs are from \citet{2019ApJ...884L...7H}. Sample A comprises objects with high probability of being identified as LBVs, which are the focus of our study, while sample B includes objects with a lower probability.
}

\label{fig:color}
\end{figure}

It is well established that both LBVs and another type of massive star-B[e]SGs are located in the upper left region of the H-R diagram. This is due to the variability of B[e]SGs and their similarities to LBVs in terms of color and spectral type \citep{1996A&A...315..510Z,2013A&A...560A..10C}. According to previous studies \citep{2013A&A...558A..17O,2014ApJ...780L..10K,2017ApJ...844...40H}, these two types differ in their infrared colors. To further determine whether these objects can be classified as cLBVs, a near-infrared color-color diagram is presented in Figure \ref{fig:color}. However, due to the lack of comprehensive IR data for J004253 and J013422, and the presence of contamination source within 3 arcseconds for J013401, we have excluded these sources from our color analysis. As a result, color data are available for only five of eight cLBVs (Sample A), and an additional 16 cLBVs (Sample B) are also included in the diagram for comparison.

As shown in Figure \ref{fig:color}, LBVs and B[e]SGs are clearly separated in the \textit{J}\,-\,\textit{H} versus \textit{H}\,-\,$\mathrm{\textit{K}_s}$ color-color diagram. The candidates from both Sample A and Sample B, with the exception of the binary system J013339, are well positioned near the known Galactic and M31 LBVs. Detailed information for both groups are listed in Table \ref{tab:color} and Table \ref{tab:lowp}.

\begin{table*}

\centering
\caption{Color Information and Stellar parameters of the eight cLBVs (Sample A).}
\setlength{\tabcolsep}{3pt}   % ← 缩小列间距
\begin{tabular}{cccccccccccc} 
\noalign{\smallskip}
\noalign{\smallskip}
         \hline 
\noalign{\smallskip}
 Index&$\mathrm{Ra^{\circ}}$&$\mathrm{Dec^{\circ}}$&\textit{H}\,-\,$\mathrm{\textit{K}_s}$ &\textit{J}\,-\,\textit{H}& $\mathrm{\textit{T}_{SED}}$ (K) & \textit{L} ($\mathrm{\textit{L}_{\odot}}$)
 & $\mathrm{\textit{T}_{SpecFit}}$ (K)& Date & Log\,$g$ & [Fe/H]\\
         \hline
            \noalign{\smallskip}
            \multicolumn{11}{c}{\textbf{J004051}}\\
               1&10.21493&40.55084&--0.159&0.794& 9000 &$4.18\times \mathrm{10^{5}}$&9249 &2011-10-24&1.21&0.50\\
            &&&&&&& 9024.9 &2011-11-17&1.17&0.92 \\
                &&&&&&&10226.8 &2015-09-20&1.40&0.79\\

  \multicolumn{11}{c}{\textbf{J004253}}\\
2&10.72255&41.45014& - &- &12500&$4.04\times \mathrm{10^{5}}$ &
% 142&
12409 &2011-10-28&1.69&0.24\\
 \multicolumn{11}{c}{\textbf{J013339}}\\

3&23.41451&30.76126& 1.878&--0.365&14500  &  $5.26\times \mathrm{10^{5}}$ 
&11374 &2018-10-09&1.56 &0.42\\
\multicolumn{11}{c}{\textbf{J013340}}\\
      4&23.41689 &30.47937&0.082 &--0.022&11400& $12.26\times \mathrm{10^{5}}$ 
      &11278& 2011-12-11&1.53&-0.01\\
      \multicolumn{11}{c}{\textbf{J013401}}\\
       5&23.50455 &30.60523& - &- & - &  - 
       &13793 &2018-10-09&2.04&0.35\\
          \multicolumn{11}{c}{\textbf{J013411}}\\
      6&23.54718 &30.7754&--0.13 &0.311&7000&   $8.91\times \mathrm{10^{5}}$

      &6838 &2011-12-11&0.49 &0.34\\            
                  \multicolumn{11}{c}{\textbf{J013420}}\\
                  7& 23.58714&30.51101&0.04&0.102&7800&  $6.07\times \mathrm{10^{5}}$&7386.4&2014-11-16&0.81&0.90\\
      &&&&&&& 7698 &2018-10-09&0.83 &0.81 \\
                 \multicolumn{11}{c}{\textbf{J013422}}\\
                 8&23.59583&30.73634& - &- &11200 & $ 4.71\times \mathrm{10^{5}}$&
       11404.1&2010-10-03&1.58&0.50\\
    &&&&&&&  11419& 2011-12-25&1.58&0.33\\
         \hline
    \end{tabular}
    \label{tab:color}
\begin{description}
  \item[$T_{SED}$:] temperature obtained from SED fitting.
  \item[$T_{SpecFit}$:] temperature obtained from spectral modeling.
\end{description}
\end{table*}

\subsection{Results}
The criteria that we classify a star as LBV mainly include several aspects: variability of light curves and spectra, existence of certain emission lines, such as He I and Balmer lines with P Cygni profile, no [O I], [Ca II] emission lines, the position in the near-infrared color-color diagram, the low outflow velocity and no evidence of warm dust in their SEDs. In the following study, we will analyze each individual in detail to confirm whether they belong to LBVs or cLBVs in terms of the above conditions.

\label{sec:individual}
\subsubsection{J004051.59+403303.0}

This star has been investigated by many previous studies. Its brightness was measured as \textit{V}\,=\,17.43 $\pm$ 0.18\,mag in 1963 \citep{1988A&AS...76...65B} and 17.33\,mag in 1990 \citep{1992A&AS...96..379M}. \citet{2015MNRAS.447.2459S} provided  \textit{V}\,=\,16.99 $\pm$ 0.05\,mag consistent with \cite{2006AJ....131.2478M}. There is a tendency that the brightness is gradually increasing. Although J004051 has previously been classified as a LBV in M31 \citep{2006AJ....131.2478M,2015MNRAS.447.2459S,2018RNAAS...2..121R}, we here present more new evidences. Figure \ref{figure:J004051.58+403303.0_lc} displays its photometric evolution, revealing a slight variation from around 2000 to 2024. The average photometric variability is approximately 1.8\,mag in the \textit{V} band.

In terms of spectroscopy, new LAMOST spectrum shows similarities to those mentioned in \citet{2015MNRAS.447.2459S} and \citet{2016ApJ...825...50G}, such as He I and Fe II absorption lines. The comparison of the spectra is plotted in Figure \ref{figure:J004051.58+403303.0}. The spectra display prominent Balmer emission lines (e.g., H$\alpha$, H$\beta$, H$\delta$, and H$\gamma$) with P Cygni profiles. Notably, many emission lines exhibited obvious changes over time, including H$\alpha$, H$\beta$, H$\delta$, H$\gamma$, He I, and Fe II. The intensity of H$\gamma$ was the strongest on 12 October 2013, exhibiting a conspicuous P Cygni profile. Then it weakened substantially by 2015. In particular, H$\beta$ follows a weak-strong-weak evolutionary pattern across the first three spectra, peaking on 12 October 2013 before declining thereafter. Absorption line [Fe II] 5154.6\,\AA\, and emission line Fe II 5235.1\,\AA\, were enhanced a lot in 2015.
% \textbf{Specifically, [Fe II] 5154.6 and Fe II 5235.1   lines have changed significantly relative to the earlier spectrum. change from emission to absorption?} 
Table~\ref{tab:vterm} additionally lists the outflow velocities independently measured from the P Cygni absorption minima. The velocity of H$\alpha$ is about --250\,km $\mathrm{s^{-1}}$, while the higher-order Balmer lines yield lower velocities ($\leq$ --180 km\,$\mathrm{s^{-1}}$). The velocity range aligns with that reported for confirmed LBVs in \citet{2014ApJ...790...48H}. According to Table \ref{tab:color} and Figure \ref{figure:J004051.58+403303.0_lc}, the effective temperature derived from three different epochs indicates that the temperature in 2011 was higher than that in 2015, suggesting that the source was likely in a minor outburst phase in 2011, while it was in a quiescent state in 2015. In Figure \ref{figure:J004051.58+403303.0_sed_specfit} (a), the existence of IR excess indicates possibility of a companion. We fitted the IR excess by adding an extra blackbody component with a temperature of 200\,K. It suggests that no warm dust surrounds the star.
The characteristics of J004051 are fully consistent with the criteria of LBV. This gives us confidence of the criteria which are used to classify LBVs.

\subsubsection{J004253.41+412700.5} 
According to \citet{2022ApJ...932...29L}, J004253 was classified as a blue supergiant candidate, with an effective temperature of 10,650 K. In Table \ref{tab:color}, higher effective temperatures ($\sim$\,12400\,K) are measured from LAMOST spectrum and SED fitting. This could be attributed to the different methods adopted. In this work, the temperature is measured by fitting the whole spectrum, whereas the lower value reported by \cite{2022ApJ...932...29L} is based on fits to the Balmer lines alone. By examining images from Pan-STARRS and Gaia, it is evident that multiple sources in the vicinity of J004253 could potentially contaminate its photometric measurements. To improve the precision of its light curve, we have excluded data from CSS and ASAS-SN because of their lower spatial resolution. The refined result is shown in Figure \ref{figure:J004253.41+412700.5_lc}, and the photometric variation is found to be approximately 0.27\,mag. The SED shown in Figure~\ref{figure:J004253.41+412700.5_sed_specfit}(a) suggests an absence of warm dust around this source.

Figure~\ref{figure:J004253.41+412700.5} shows pronounced Balmer lines in all spectra. Multi-epoch observations reveal clear temporal evolution in both the strength and profiles of these lines. The trends of variation for both He I 4026.63 and H$\delta$ are consistent, following a emission-absorption-emission pattern. More importantly, the He I,  H$\beta$, H$\delta$ and H$\gamma$ lines all displayed P Cygni profiles and their intensities showed marked variability on multi-year timescales. Moreover, the star exhibited low outflow velocities ($\leq$ --150\,km $\mathrm{s^{-1}}$) in all Balmer lines listed in Table~\ref{tab:vterm}. In particular, the He I line yields an unexpected small value of --51.2\,km $\mathrm{s^{-1}}$. Even though the photometric variability is weak, the absence of warm dust, low outflow speed, and marked spectroscopic variability secure its designation as a cLBV.

\subsubsection{J013339.48+304540.5}

J013339 was classified as a blue supergiant in M31 by \cite{1993ApJS...89...85I}.  Figure \ref{figure:J013339.48+304540.5_lc} shows its light curve, covering over 20 yrs. In \textit{V} band, the source exhibits a change of approximately 0.91\,mag over the entire time span. \citet{2016AJ....152...62M} classified this source as a cLBV with \textit{V}\,=\,17.5\,mag, \textit{B}\,-\,\textit{V}\,=\,0.064\,mag. According to \citet{2006AJ....131.2478M}, J013339 is a binary system with a composite spectral type of B0.5Ia $+$ WNE. Moreover, we discovers a period of $\sim$\,33.1126\,days for the first time, using ZTF photometric data (Liu et al., in prep.). As shown in Figure \ref{figure:J013339.48+304540.5}, the Balmer emission lines appeared relatively stronger than other lines, underwent evident change and reached its maximum intensity in 2018. Additionally, spectral variability is also observed in the He I, [O III], [N II], and [S II] lines. Among these lines, He I exhibited its highest intensity in 2013 relative to all other epochs, while the [O III], [N II], and [S II] features are likely shaped by nebular emission from the surrounding H II region \citep{2014ApJ...790...48H}. Like J004253, the outflow velocities cluster around --100 km\,$\mathrm{s^{-1}}$, with the He I line again giving the smallest value compared to other lines. As in the case of previous object, the SED of J013339 also shows an IR excess and no evidence of warm dust. Similarly, we attribute such an excess to the effectivity of H II region \citep{2014ApJ...790...48H}.

In previous work, \citet{2013A&A...560A..10C} discussed the spectral variability of LHA 115-S 18 (S18) in B[e]SGs, evaluating whether S18 belongs to a binary or LBV classification. J013339 appears to be a similar case: a binary system exhibiting LBV-like spectral variability. Taking all available physical parameters into account, we therefore classify this source as a candidate LBV.

\subsubsection{J013340.05+302845.7}

The \textit{V}-band magnitude of J013340 is 16.047\,mag and its \textit{B}\,-\,\textit{V} is 0.118\,mag, reported by \citet{2006AJ....131.2478M}. 0Recently, \citet{2022ApJ...932...29L} classified it as a blue supergiant candidate in M33. As shown in Figure \ref{figure:J013340.05+302845.7_lc}, J013340 exhibits almost no considerable changes in its light curve. Using the same calculation method mentioned above, the measured $\Delta$\textit{V} is 0.86\,mag. Combining multi-band photometry, the SED was fitted in Figure \ref{figure:J013340.05+302845.7_sed_specfit} (a) with $\mathrm{\textit{T}_{SED}}$\,=\,11,400\,K and no warm dust.

Two spectra were obtained from LAMOST, observed on December 11, 2011 and December 1, 2014. In Figure \ref{figure:J013340.05+302845.7}, the spectrum from 2011 displays typical LBV spectral line features: variable Balmer emission lines and He I emission lines. Due to the low SNR of the second spectrum, a detailed comparison is challenging. It is seen that the H$\alpha$ emission line evidently weakened over the three-year interval. Without prominent P Cygni profiles, there is also trouble in calculating the outflow velocity. Higher-quality spectra are needed to investigate the spectral variable nature. Consequently, there is no strong evidence supporting this star as a cLBV.

\begin{figure}[htbp]
\includegraphics[angle=0,scale=0.85]{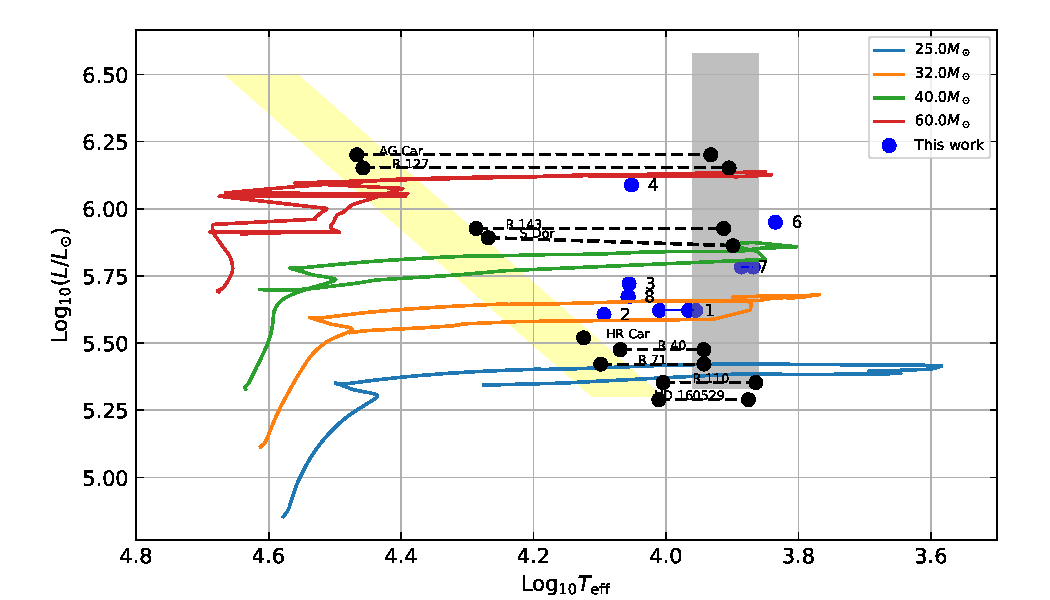}
\centering

\caption{H-R diagram for LBVs. The shaded region on the right indicates the outburst phase of LBV. The shaded area on the left depicts the diagonal S Doradus instability strip, according to \citet{1994PASP..106.1025H}. Black points correspond to established LBVs, with luminosities (\textit{L}) and effective temperatures ($\mathrm{\textit{T}_{eff}}$) sourced from \citet{1994PASP..106.1025H} and \citet{1998A&ARv...8..145D}. Blue points denote the subjects of our study, with \textit{L} values derived in Section \ref{sed} and $\mathrm{\textit{T}_{eff}}$ values derived in Section \ref{specm}. Additionally, solid lines represent the Geneva evolutionary tracks \citep{2012A&A...537A.146E} for solar metallicity and an initial rotational velocity of 40\% of the critical breakup speed. For the sake of simplicity, only the initial 209 points are included.
}

\label{fig:HR}
\end{figure}

\subsubsection{J013401.16+303618.4}

According to \citet{2017ApJ...836...64H}, J013401 is a hot supergiant with \textit{V}\,=\,18.23\,mag provided by \citet{2006AJ....131.2478M}. We find that there are two objects within 3 arcseconds in the image of Pan-STARRS and Gaia. Hence, its position in the near-infrared color-color diagram, light curve, and the SED fitting are not considered in this work. Without stellar parameters derived from the SED fitting, its location in Figure~\ref{fig:HR} remains unconstrained. With respect to spectroscopy, three spectra are presented in Figure \ref{figure:J013401.16+303618.4}. The two spectra obtained from \citet{2017ApJ...836...64H} have higher SNR than the one captured by LAMOST. The LAMOST spectrum exhibits significant numbers of artifacts, especially in the blue region.

J013401 shows narrow but intense Balmer lines (with P Cygni profiles), He I, [O III], [O I], [N II], and [S II] lines. And [O III], [O I], [N II], and [S II] lines may originate from the possible emission of surrounding H II region. Additionally, there are notable variations in the Balmer series and He I line. With the P Cygni profile of H$\gamma$ and He I lines, the outflow velocities were calculated to be --158.82\,km $\mathrm{s^{-1}}$ and --67.56\,km $\mathrm{s^{-1}}$, respectively. These features align with the spectral characteristics of LBVs, as discussed by \citet{2019MNRAS.484L..24S} and \citet{2020MNRAS.497..687S}. 

The presence of these spectral lines, spectroscopic variability, and the low outflow velocities align well with above criteria for LBVs. However, due to the absence of photometric analysis, we still keep it as a hot supergiant.

\subsubsection{J013411.32+304631.4}

The brightness of this star was given in \citet{2006AJ....131.2478M}, as \textit{V}\,=\,15.996\,mag and \textit{B}\,-\,\textit{V}\,=\,0.332\,mag. Moreover, \citet{2012ApJ...750...97D} studied it as a yellow supergiant, deriving $T_{\rm eff}$ of 6982.32\,K. After searching for photometry data of J013411 in the CSS and ASAS-SN surveys within 3 arcseconds, no objects were found. Upon increasing the search radius to 6 arcseconds, the obtained data were evidently contaminated by surrounding sources. Consequently, the light curve is presented without the data from these two surveys. This leads to the presentation in Figure \ref{figure:J013411.32+304631.4_lc}. Overall, the variation of photometry is 0.14\,mag. The SED of the subject was constructed and an IR excess was fitted by adding an extra blackbody component in Figure \ref{figure:J013411.32+304631.4_sed_specfit} (a), obtaining effective temperature of 7000\,K, which is in agreement with 6982.32\,K within uncertainty. And the temperature obtained from fitting the excess is 200\,K, it is definitely not enough for warm dust. In Figure \ref{fig:HR}, it is seen that the star is located on the right side with lower temperature compared to other objects. However, based on \citet{1994PASP..106.1025H}, in the eruptive state, it is reasonable for LBVs to have cooler temperature around 7000-8000\,K. 

Two LAMOST spectra obtained within a short time interval are presented in Figure \ref{figure:J013411.32+304631.4}. The spectra show narrow Balmer lines, Fe II, He I lines, [N II], and [S II] lines. The lines [N II], [S II] may be from surrounding nebula. Variability of spectral lines are observed in He I 4026.63, H$\gamma$ and H$\delta$ lines, as well as He I 4389.16 and [S II], all of which show noticeable line broadening. In the red spectral region, both the H$\alpha$ and the [S II] line at 6718.29\,\AA\,show changes. Their emission intensity decreases, while only [S II] line broadens. To further confirm whether this star meets our classification criteria, we measure the outflow velocity to be 242.98\,km $\mathrm{s^{-1}}$ from the inverse P Cygni profile of H$\alpha$. According to \citet{1990A&A...235..340W}, such a distinguished inverse P Cygni profile is indicative of infalling material. Eventually, considering its position in the H-R diagram, it is not enough to claim that J013411 is a cLBV.

\subsubsection{J013420.91+303039.6}

The star was regarded as a yellow supergiant by \citet{2012ApJ...750...97D} and \citet{2016ApJ...825...50G}, and its spectral type corresponds to A8Ia in \citet{2017A&A...601A..76K}. \citet{2012ApJ...750...97D} and \citet{2016ApJ...825...50G} independently derived effective temperature of 7894.28\,K and 8550.67\,K, respectively. In our study, photometric variability is observed, with amplitude of 0.32\,mag in the \textit{V} band. As depicted in Figure \ref{figure:J013420.91+303039.6_lc}, photometric data from ASAS-SN has been excluded due to severe blending that affects the quality of the light curve. The best-fitting SED model is shown in Figure \ref{figure:J013420.91+303039.6_sed_specfit} (a). Our SED fitting suggests that effective temperature is 7800\,K, consistent with that of \citet{2012ApJ...750...97D}. And the IR excess is fitted with a temperature of 150\,K. Given this temperature, we can infer that no warm dust is present around the star. 

In Figure \ref{figure:J013420.91+303039.6}, five spectra are compared, revealing a significant feature that H$\delta$, H$\gamma$ and H$\beta$ all transitioned from emission to absorption between 2012 and 2014. In terms of line intensity changes, the obvious ones include He I, Mg II, [O III], Fe II, [Fe II], Si II, [N II], H$\alpha$, and [S II]. The attractive aspect is that the intensity of the [N II], [S II], and H$\alpha$ lines follows a weak-strong-weak pattern during evolution, and they have conspicuously broadened in 2012. Overall, the spectral variations within a short timescale are quite pronounced. The P Cygni absorption minimum of H$\gamma$ yields a low outflow velocity of --115.00\,km $\mathrm{s^{-1}}$. Spectral modeling are separately performed  for the two spectra observed on 2014 November 16 and 2018 October 09. The derived effective temperatures are in close agreement (see star 7 in Figure \ref{fig:HR}), in line with the absence of noticeable variability in the light curve. Despite the slight photometric variability, the combination of the object's location in the color-color diagram, absence of warm dust, spectroscopic signatures, and low outflow velocity collectively yield a high-confidence classification of cLBV.

\subsubsection{J013422.87+304410.8}
The object J013422 was determined as a cLBV by \citet{2007AJ....134.2474M} with \textit{V}\,=\,17.22\,mag and \textit{B}\,-\,\textit{V}\,=\,0.07\,mag. The light curve of this star is shown in Figure \ref{figure:J013422.87+304410.8_lc}, with an average brightness variation of $\Delta$\textit{V} $\approx$\,0.89\,mag. It is evident that the source appears brighter from 2006 to 2024 compared with other epochs. For the discrepancy between the ASAS-SN and ZTF light curves, it can be attributed to systematic effects in the ASAS-SN data. In particular, the large pixel scale of ASAS-SN ($\sim 8^{\prime\prime}$
per pixel) makes the photometry susceptible to source blending and contamination from nearby objects in crowded regions. This blending effect can lead to systematically overestimated brightness. As pointed out by \citet{2019MNRAS.485..961J}, ASAS-SN \textit{V}-band photometry in high-density fields such as the LMC shows a significant systematic offset, appearing brighter when compared with APASS measurements.

By comparing the four spectra, spectral variations are also observed in Figure \ref{figure:J013422.87+304410.8}. The typical lines, including H$\delta$, H$\gamma$ (with P Cygni profile), H$\beta$, H$\alpha$, He I, Si II, and Fe II have undergone intensity changes. Regarding the status change in the spectral lines, near 6400\,\AA, a [O I] line which may be from surrounding nebula changed from absorption to emission. The outflow velocity measured in P Cygni profile of H$\alpha$ was --107.85\,km $\mathrm{s^{-1}}$. As shown in Table~\ref{tab:color} and Figure \ref{fig:HR},, the effective temperatures measured at different epochs are mutually consistent, implying that the star has remained in the same evolutionary phase during the period from 2010 to 2011. The phase was located between S Doradus instability strip and outburst phase. Due to the absence of $\mathrm{\textit{JHK}_s}$ photometry, it is difficult to identify where the star is located in Figure \ref{fig:color}. J013422 lacks definitive evidence for a LBV classification.

\subsection{Other candidates}
Upon checking the spectra of additional sixteen sources, we observed prominent Balmer emission lines. Consequently, we collected their photometric color-color data and plotted them in Figure \ref{fig:color}. We find that they overlap with LBVs rather than B[e]SGs. Considering the lack of more available information, they are just labeled as suspected cLBVs, detailed in Table \ref{tab:lowp}.

\begin{table}
\caption{Suspected cLBVs Catalog (Sample B).}
\centering
\label{tab:lowp}

\begin{tabular}{ccccc} 
\noalign{\smallskip}
         \hline
\noalign{\smallskip}

 Name&$\mathrm{Ra^{\circ}}$&$\mathrm{Dec^{\circ}}$&\textit{H}\,-\,$\mathrm{\textit{K}_s}$ &\textit{J}\,-\,\textit{H}\\
         \hline
            \noalign{\smallskip}
                  J004033.90+403047.1&10.14126& 40.51309& -0.070&-0.072 \\
                 J004109.46+404900.1&10.28943& 40.81670&-0.463&0.764 \\
                 J004339.36+411008.6&10.91401 &41.169080 &0.052&0.221 \\
                 J004510.03+413657.6&11.29186&41.61593&0.074&0.130\\
                 J013315.36+303722.6&23.31401 &30.62295 &0.219&0.761 \\
                 J013316.36+305320.5&23.31818 &30.88903&0.018&1.016 \\
                 J013328.69+304134.9&23.36954&30.69310&0.402&0.230 \\
                 J013333.10+302039.5&23.38793 &30.34431&0.250&0.930 \\
                 J013351.30+303855.3& 23.46376 &30.64870&0.265&0.777 \\
                 J013401.94+303713.1&23.50809 &30.62031&0.450&0.680\\
                 J013403.01+304410.6&23.51255 &30.73628&-0.135&0.272 \\
                 J013409.66+303917.6&23.54026 &30.65490&0.277&0.852 \\
                 J013415.50+303711.5&23.564333&30.619942	&0.577&1.069 \\
                 J013440.89+304619.2&23.67039 &30.77201&-0.054&0.028 \\
                 J013441.92+305610.1&23.67468 &30.93615&-0.130&0.063 \\
                 J013450.09+304704.1&23.70872 &30.78448&0.183&0.423 \\
 \noalign{\smallskip}
         \hline
            \noalign{\smallskip}
    \end{tabular}

\end{table}
\section{Discussion And Conclusions} \label{sec:conclusion}
In this study, simultaneous optical spectroscopy and photometric data analysis of stars in the galaxies M31 and M33 are presented. We determined the outflow velocities of these stars, key stellar properties through SED analysis, including temperature and luminosity, to confirm the classification of these stars. In addition, spectral modeling provides independent estimates of $\mathrm{\textit{T}_{eff}}$, Log\,$g$, and [Fe/H].

In Figure \ref{fig:HR}, we present a temperature-luminosity diagram with the evolutionary tracks of massive stars superimposed. By aligning these tracks with the Geneva evolutionary models for solar metallicity and an initial rotational velocity of 40\% of the critical speed \citep{2012A&A...537A.146E}, the initial mass of these objects are estimated to be in the range of approximately 32 to 60 $\mathrm{\textit{M}_{\odot}}$.  Owing to difficulties in constructing its SED, the luminosity of J013401 could not be reliably determined, so its position in this diagram remains undetermined, whereas the cooler object J013411 lies at the far right. Other stars fall in the location between quiescent phase and outburst phase for LBVs.

Among the eight in-depth analyzed objects with high probability, two stars have been studied in M31, while additional six stars are locate in M33. Regarding J004051, although it was previously identified as an LBV, this study provides an extensive array of photometric and spectroscopic data that further substantiates its physical nature. Individual \textit{V}-band light curves for these stars are constructed covering more than ten years, revealing relatively small variations in brightness. Despite these subtle changes, they are consistent with one of the theoretical variability states of LBVs. Upon comparing the spectra of the cLBVs, we found that they closely match the spectra of known LBVs, characterized by broad hydrogen lines, He I lines, and an abundance of Fe II and [Fe II] emission lines. Notably, there are discernible spectral variations among their multi-epoch spectra. From the \textit{J}\,-\,\textit{H} versus  \textit{H}\,-\,$\mathrm{\textit{K}_s}$ color-color diagram, it is obvious that our studied objects, with the exception of the binary system J013339 and those stars without near infrared data, are situated in the typical region associated with LBVs rather than that of B[e]SGs. Eventually, combining all the available up to date evidences for our eight cLBVs, four stars including J004253, J013339, J013420, and J013422 were incorporated into cLBVs.

For the remaining sixteen objects which have limited data, only preliminary spectral analysis are conducted. Conspicuous Balmer emission lines are noticed, suggesting that these objects could be potential cLBVs. In addition, they are also located in the typical region associated with LBVs in the near infrared color-color diagram. However, a definitive classification will require additional observational data in the future.

\begin{acknowledgements}

This work was supported by Young Scholar Program of Beijing Academy of Science and Technology (24CE-YS-08, 25CE-YS-02), the Popular Science Project (24CD012) of Beijing Academy of Science and Technology, National Natural Science Foundation of China (NSFC no. 12033003, 12203006), Beijing Natural Science Foundation (no. 1242016) and the Innovation Project of Beijing Academy of Science and Technology (11000023T000002062763-23CB059). 

The spectroscopic data presented in this study were acquired using the Guoshoujing Telescope, also known as the Large Sky Area Multi-Object Fiber Spectroscopic Telescope (LAMOST). This telescope is financially supported by the National Development and Reform Commission and is operated and overseen by the National Astronomical Observatories, which falls under the Chinese Academy of Sciences. 

SIMBAD, GAIA and ZTF provide the abundant photometric data. SIMBAD database \citep{2000A&AS..143....9W} is operated at CDS, Strasbourg, France. GAIA data are being processed by the Gaia Data Processing and Analysis Consortium (DPAC). Funding for the DPAC is provided by national institutions, in particular the institutions participating in the Gaia MultiLateral Agreement (MLA). ZTF, is supported by the National Science Foundation under Grants no. AST-1440341 and AST-2034437 and a collaboration including current partners Caltech, IPAC, the Oskar Klein Center at Stockholm University, the University of Maryland, University of California, Berkeley, the University of Wisconsin at Milwaukee, University of Warwick, Ruhr University, Cornell University, Northwestern University and Drexel University. Operations are conducted by COO, IPAC, and UW. This publication makes use of VOSA, developed under the Spanish Virtual Observatory (\href{https://svo.cab.inta-csic.es}{https://svo.cab.inta-csic.es}) project funded by MCIN/AEI/10.13039/501100011033/ through grant PID2020-112949GB-I00.
VOSA has been partially updated by using funding from the European Union's Horizon 2020 Research and Innovation Programme, under Grant Agreement $n^{o}$ 776403 (EXOPLANETS-A).

\end{acknowledgements}

\begin{appendix}                 %%appendicial material is supported

\section{Photometry Figures }
In the following graphs, the long duration of light curves of eight high-probability cLBVs are provided.
 \label{sec:phot}
\setcounter{figure}{0}
\renewcommand{\thefigure}{A\arabic{figure}}

\begin{figure}[htbp]
\begin{minipage}[t]{0.6\linewidth}
\centering
\includegraphics[scale=0.5]{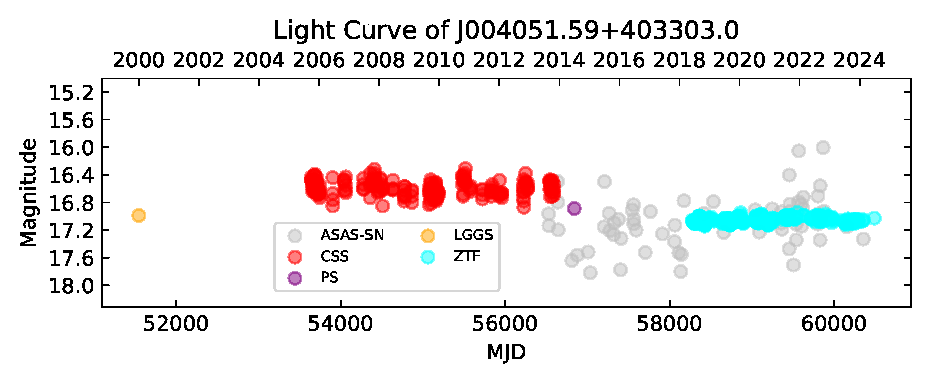}
\caption{}
\label{figure:J004051.58+403303.0_lc}
\end{minipage}%
\begin{minipage}[t]{0.5\linewidth}
\centering
\includegraphics[scale=0.5]{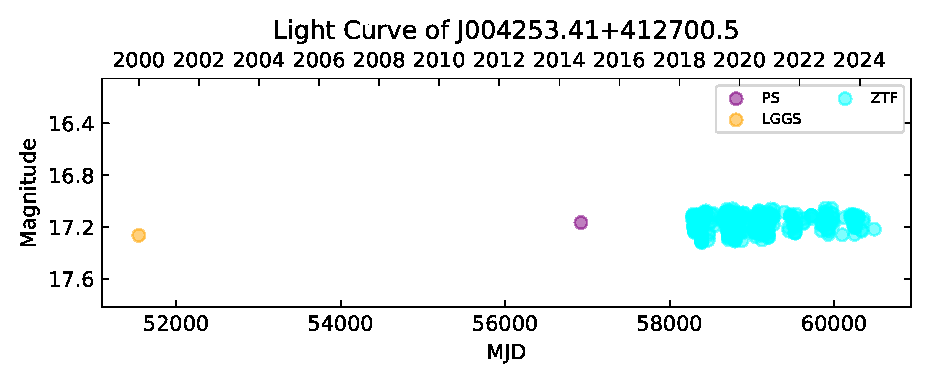}
\caption{}
\label{figure:J004253.41+412700.5_lc}
\end{minipage}
\end{figure}
\vspace{-2.5mm} % 在这里插入10mm的垂直空间

\begin{figure}[htbp]
\begin{minipage}[t]{0.6\linewidth}
\centering
\includegraphics[scale=0.5]{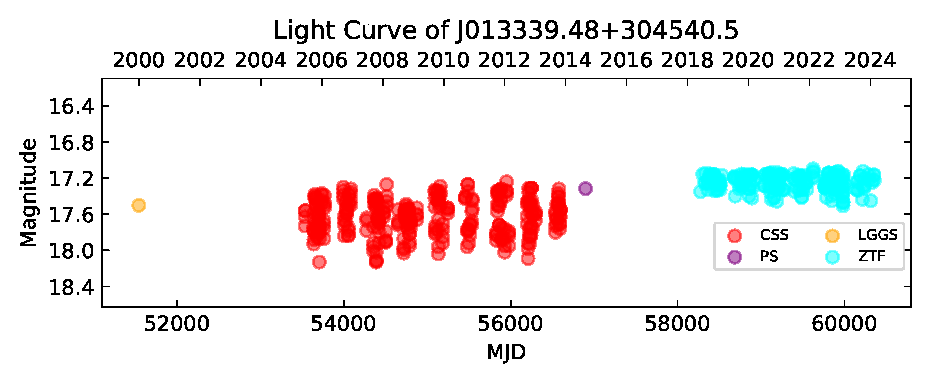}
\caption{}
\label{figure:J013339.48+304540.5_lc}
\end{minipage}%
\begin{minipage}[t]{0.5\linewidth}
\centering
\includegraphics[scale=0.5]{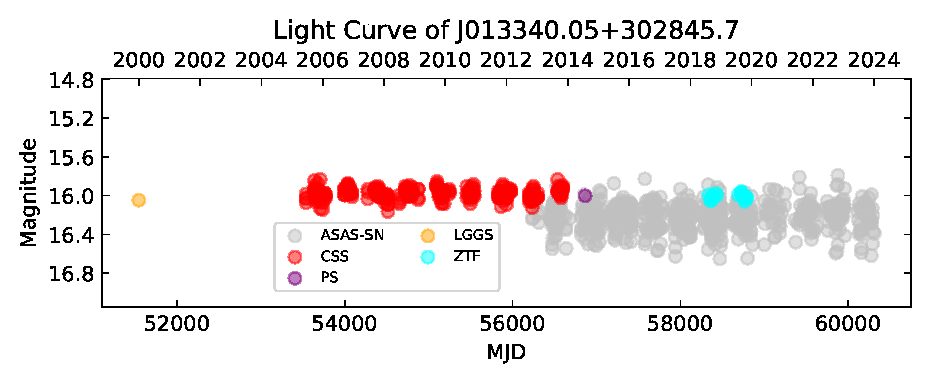}
\caption{}
\label{figure:J013340.05+302845.7_lc}
\end{minipage}
\end{figure}
\vspace{-2.5mm} % 在这里插入10mm的垂直空间

\begin{figure}[htbp]
\begin{minipage}[t]{0.6\linewidth}
\centering
\includegraphics[scale=0.5]{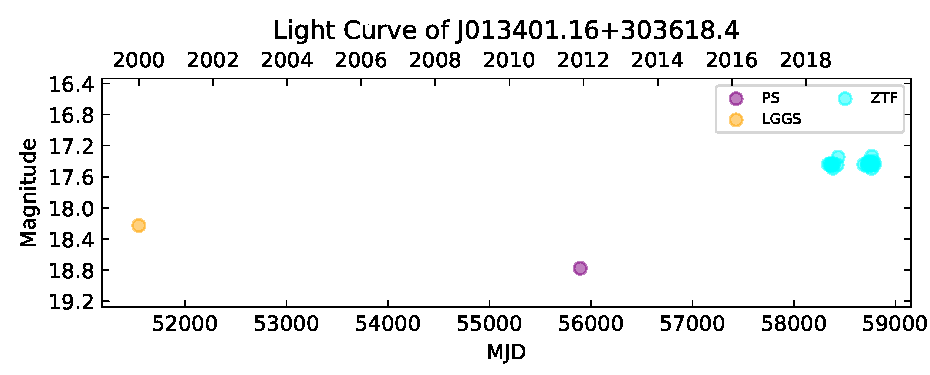}
\caption{}
\label{figure:J013401.16+303618.4_lc}
\end{minipage}%
\begin{minipage}[t]{0.5\linewidth}
\centering
\includegraphics[scale=0.5]{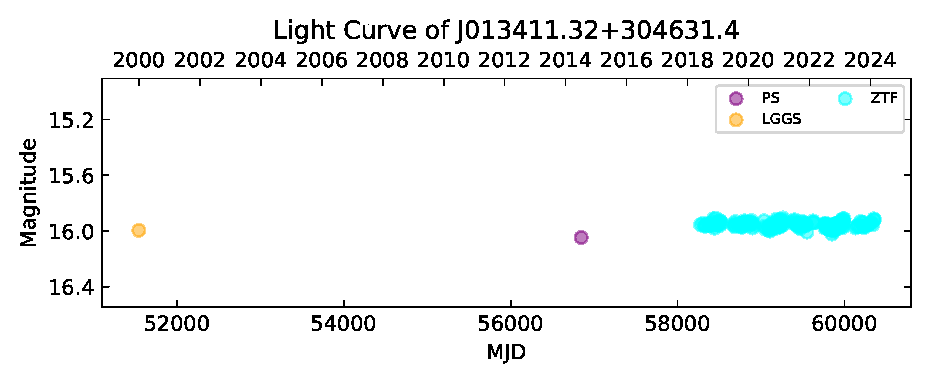}
\caption{}
\label{figure:J013411.32+304631.4_lc}
\end{minipage}
\end{figure}
\vspace{-2.5mm} % 在这里插入10mm的垂直空间
\begin{figure}[htbp]
\begin{minipage}[t]{0.6\linewidth}
\centering
\includegraphics[scale=0.5]{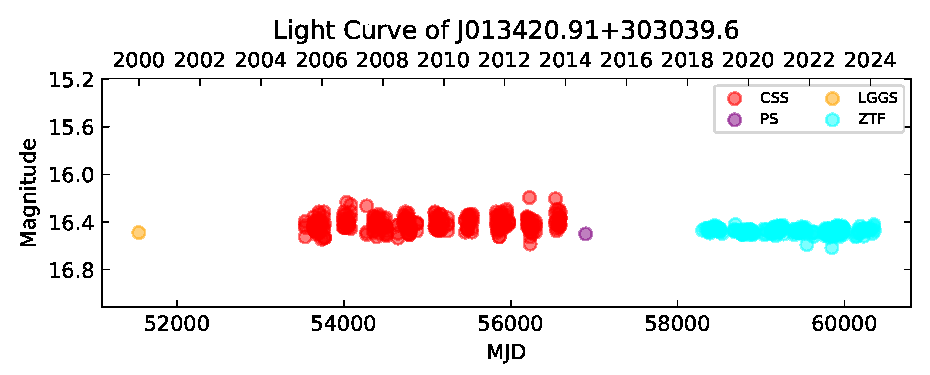}
\caption{}
\label{figure:J013420.91+303039.6_lc}
\end{minipage}%
\begin{minipage}[t]{0.5\linewidth}
\centering
\includegraphics[scale=0.5]{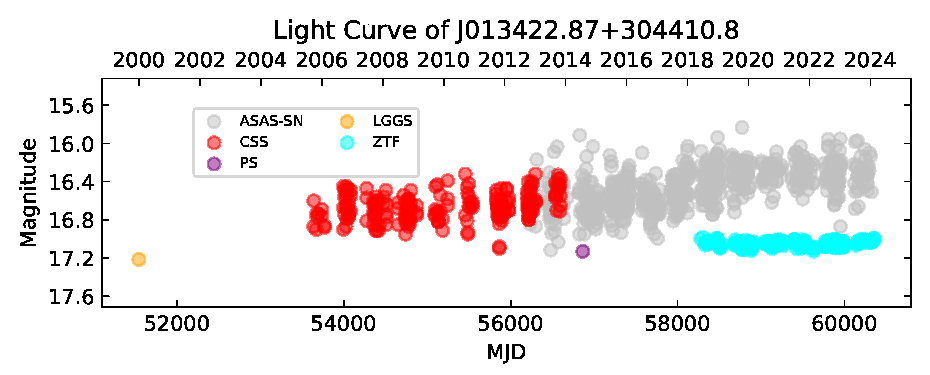}
\caption{}
\label{figure:J013422.87+304410.8_lc}
\end{minipage}
\end{figure}

\clearpage    % 强制清空附录 A 的浮动体
\section{Spectroscopy Figures }
In the following graphs, the multi-epoch spectra of eight high-probability cLBVs are provided. There are vertical lines where visible changes have occurred. For lines without textual annotations, these are unidentified lines.
 \label{sec:spectra}
\setcounter{figure}{0}
\renewcommand{\thefigure}{B\arabic{figure}}

\begin{figure}[htbp]
\begin{minipage}[t]{0.6\linewidth}
\centering
\includegraphics[scale=0.4]{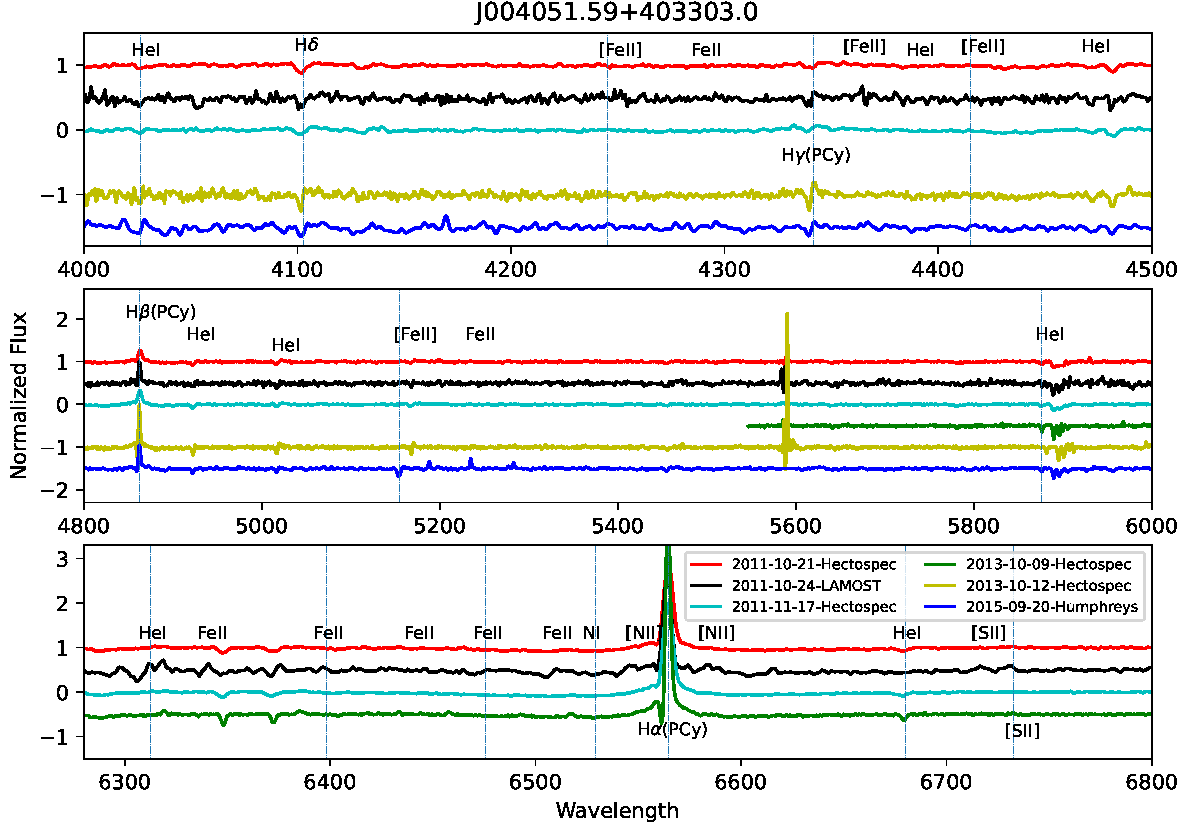}
\caption{}
\label{figure:J004051.58+403303.0}
\end{minipage}%
\begin{minipage}[t]{0.5\linewidth}
\centering
\includegraphics[scale=0.4]{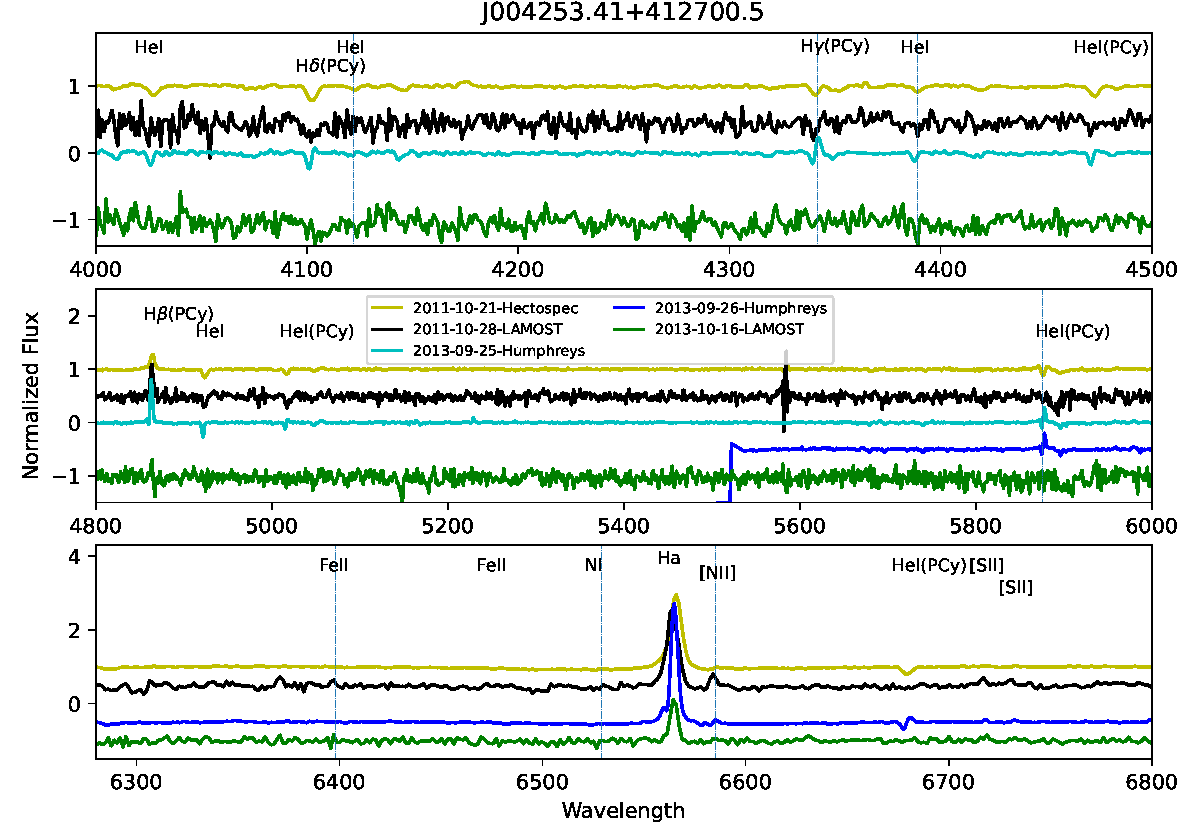}
\caption{}
\label{figure:J004253.41+412700.5}
\end{minipage}
\end{figure}

\begin{figure}[htbp]

\begin{minipage}[t]{0.6\linewidth}
\centering
\includegraphics[scale=0.4]{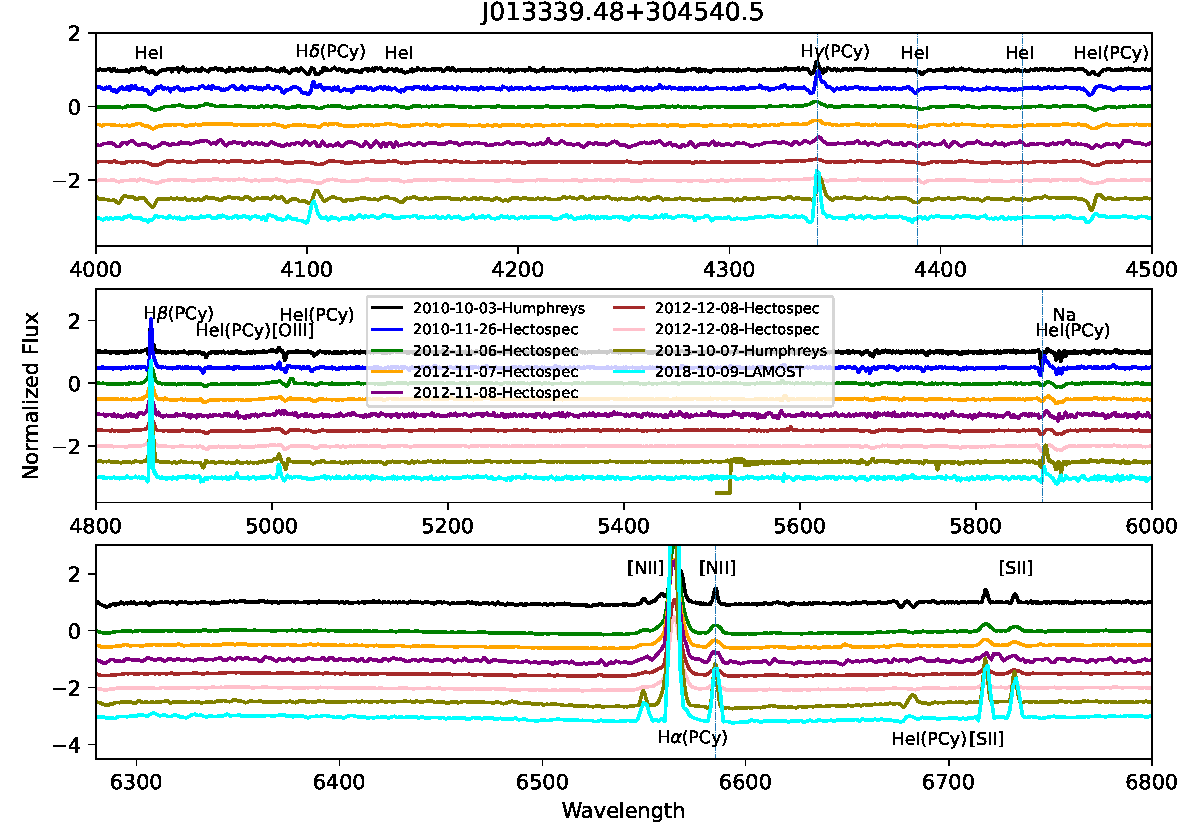}
\caption{}
\label{figure:J013339.48+304540.5}
\end{minipage}%
\begin{minipage}[t]{0.5\linewidth}
\centering
\includegraphics[scale=0.4]{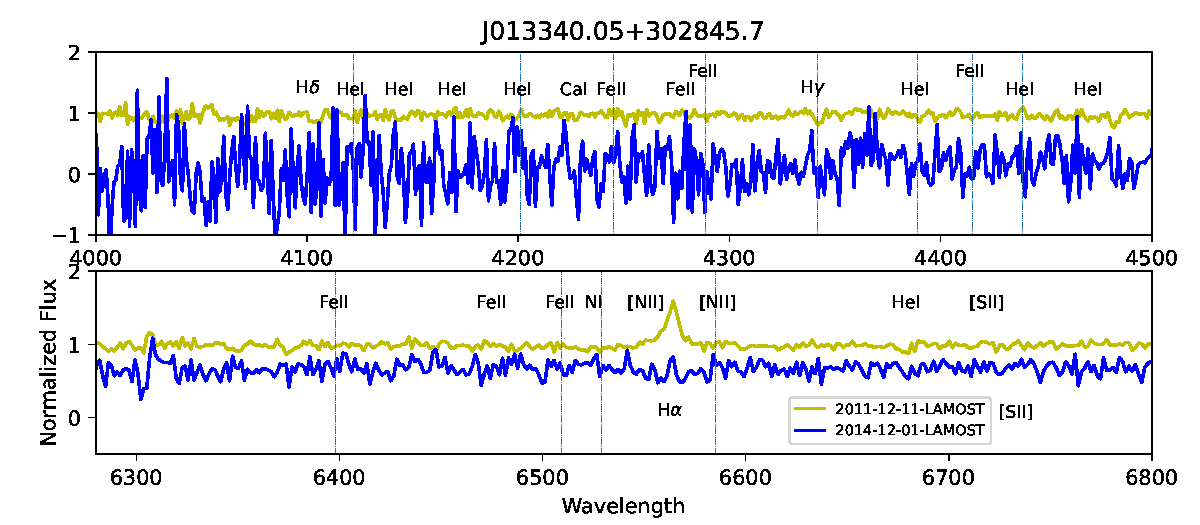}
\caption{}
\label{figure:J013340.05+302845.7}
\end{minipage}
\end{figure}

\begin{figure}[htbp]
\begin{minipage}[t]{0.6\linewidth}
\centering
\includegraphics[scale=0.4]{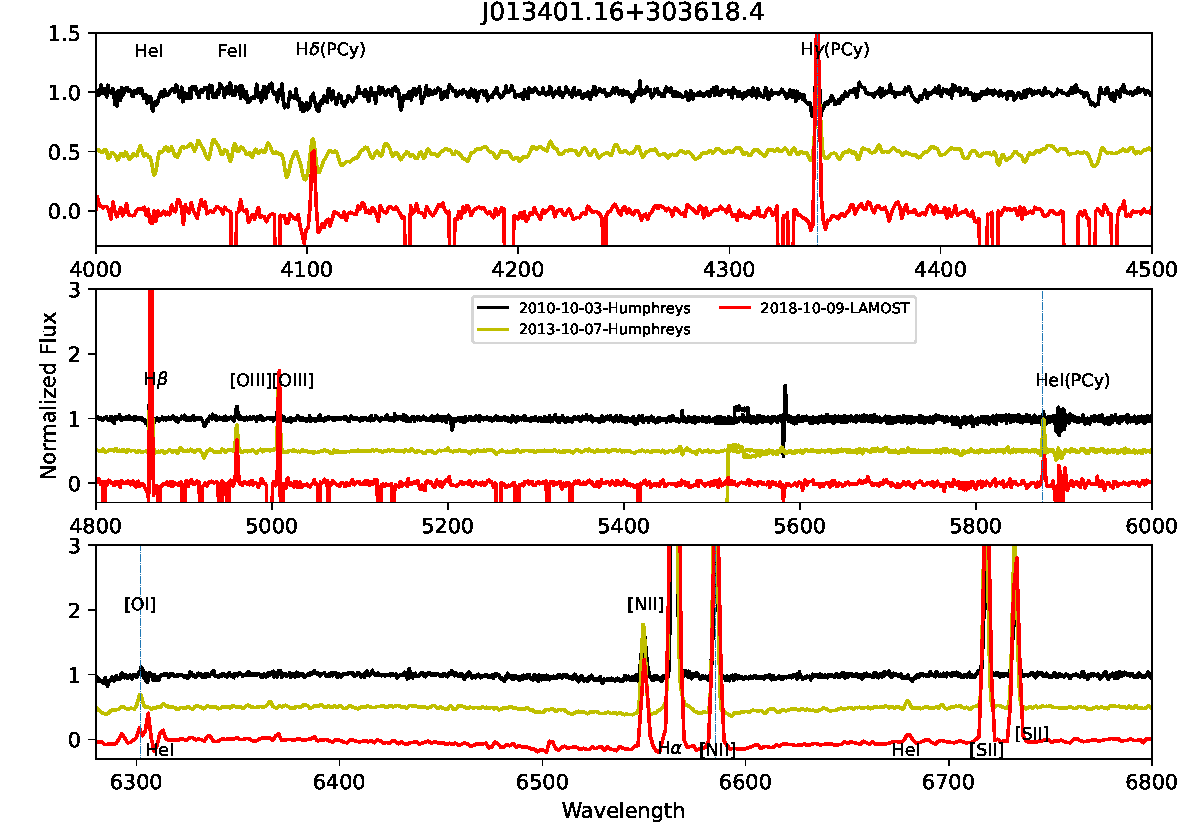}
\caption{}
\label{figure:J013401.16+303618.4}
\end{minipage}%
\begin{minipage}[t]{0.5\linewidth}
\centering
\includegraphics[scale=0.4]{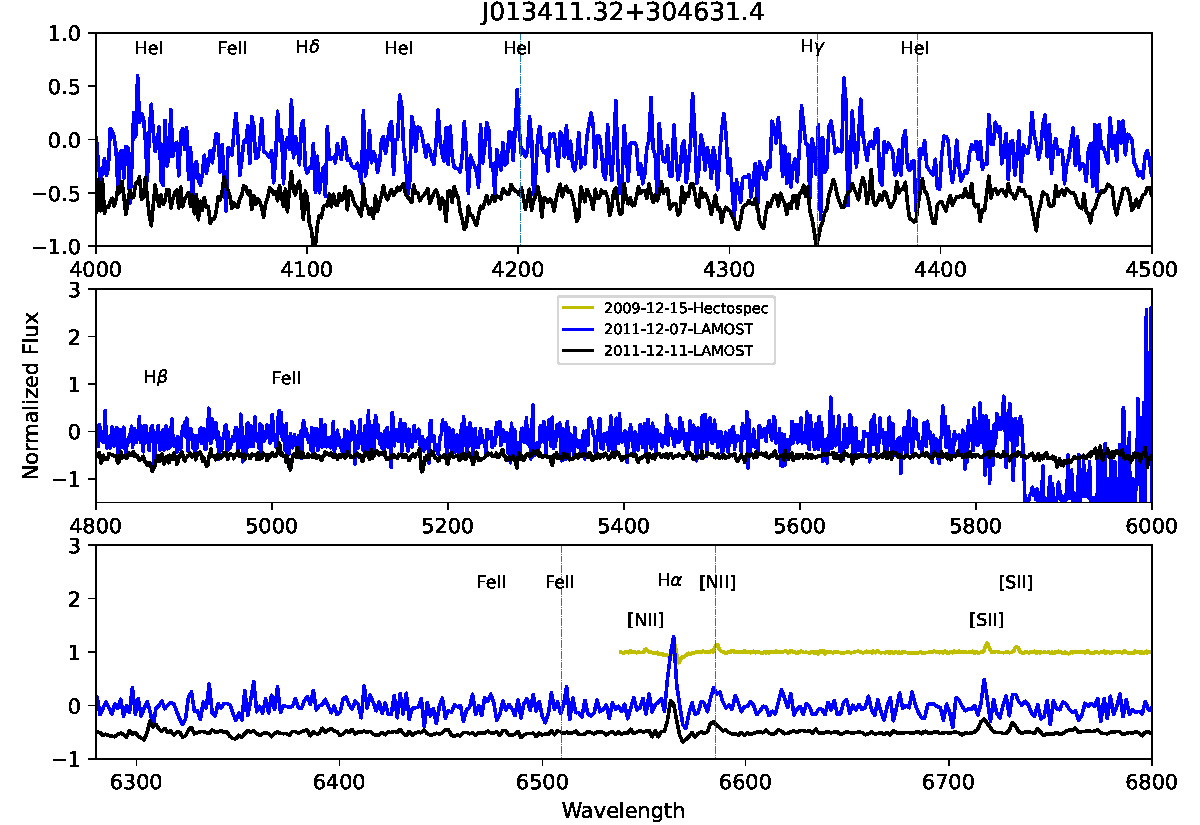}
\caption{}
\label{figure:J013411.32+304631.4}
\end{minipage}
\end{figure}

\begin{figure}[htbp]
\begin{minipage}[t]{0.6\linewidth}
\centering
\includegraphics[scale=0.4]{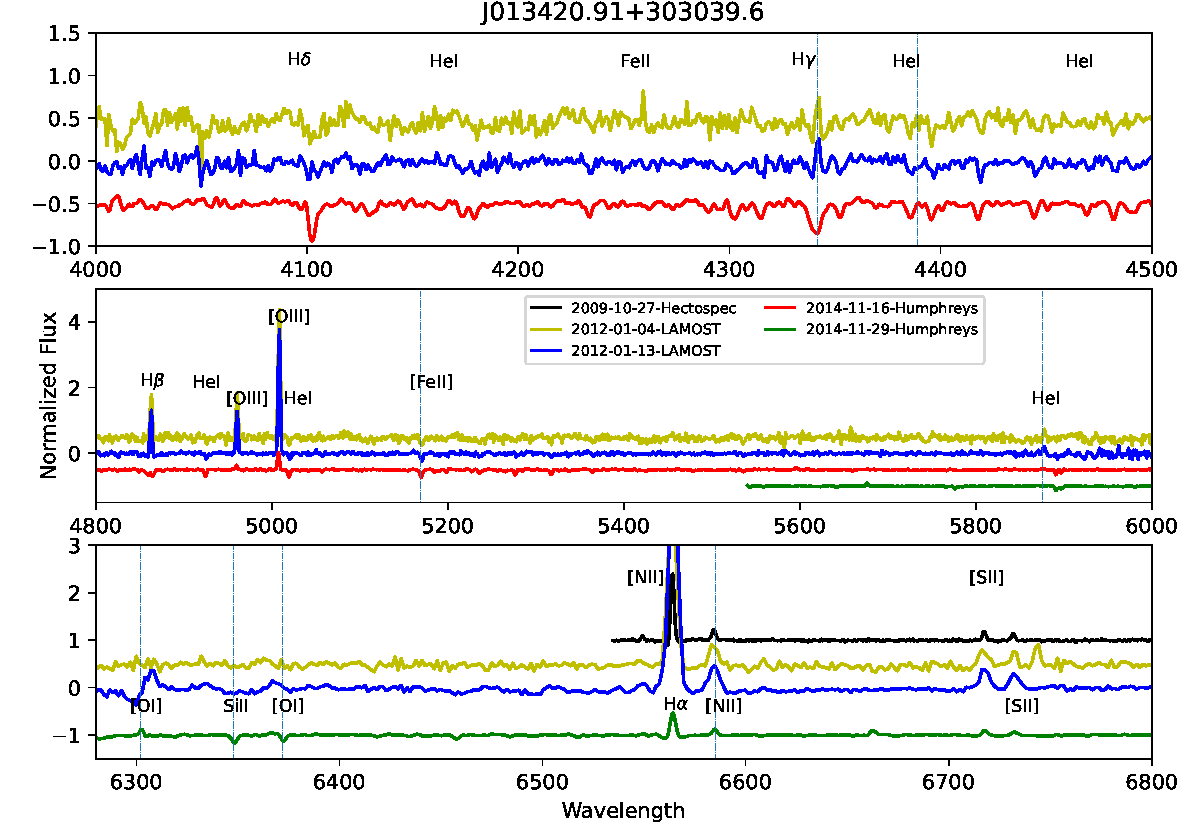}
\caption{}
\label{figure:J013420.91+303039.6}
\end{minipage}%
\begin{minipage}[t]{0.5\linewidth}
\centering
\includegraphics[scale=0.4]{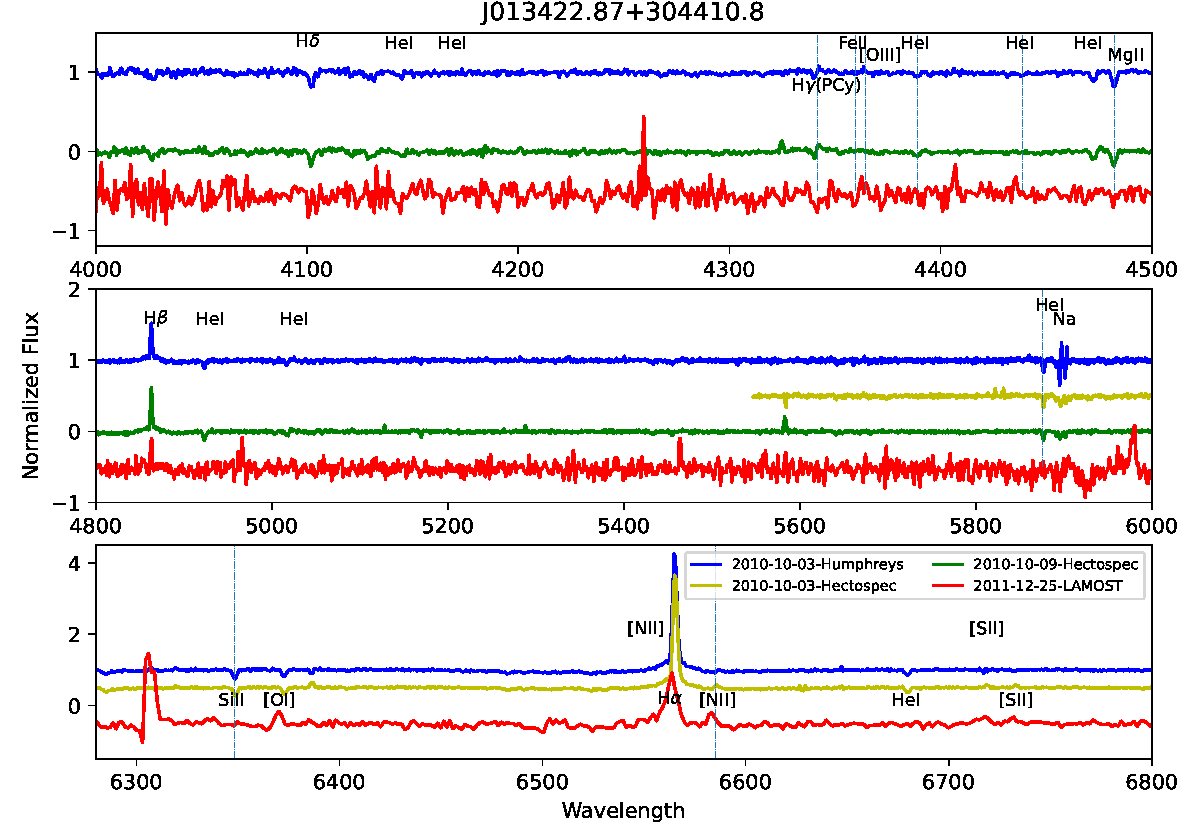}
\caption{}
\label{figure:J013422.87+304410.8}
\end{minipage}
\end{figure}

\clearpage    % 强制清空附录 A 的浮动体
\section{SED fitting and Spectral modeling Figures }
 \label{sec:specfit}
\setcounter{figure}{0}
\renewcommand{\thefigure}{C\arabic{figure}}

\begin{figure}[htbp]
\centering

\subfloat[]{\includegraphics[scale=0.4]{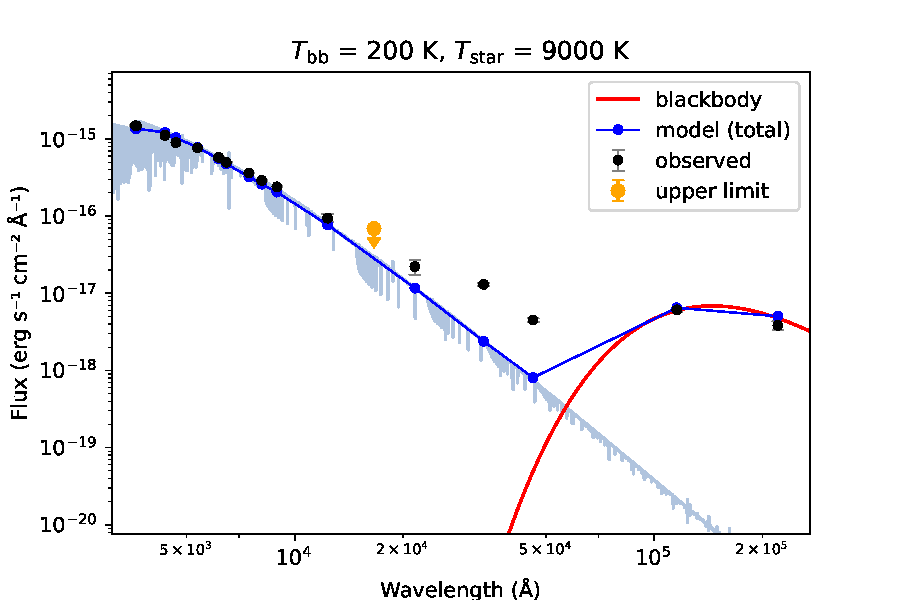}}
% \hfill
\subfloat[]{\includegraphics[scale=0.55]{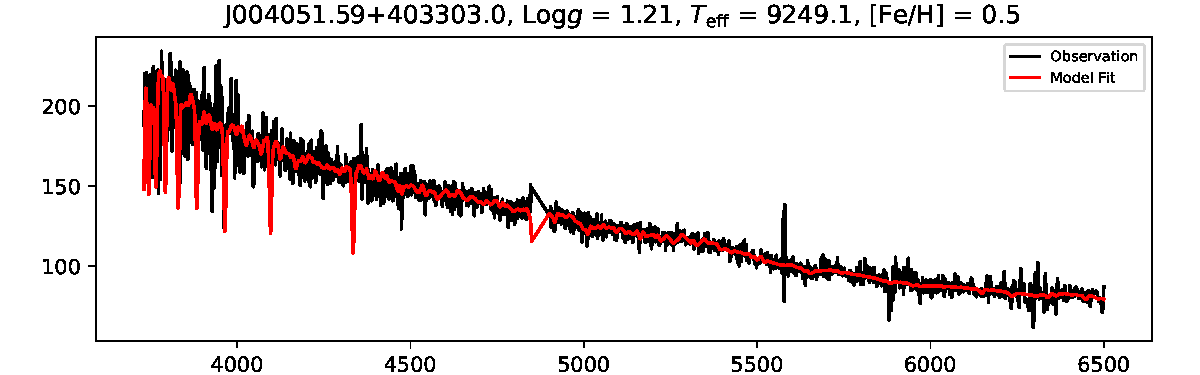}}
% \subfloat[]{\includegraphics[scale=0.4]{J004051_SED.pdf}}
% % \hfill
% \subfloat[]{\includegraphics[scale=0.55]{J004051.59+403303.0_specfit.pdf}}
\caption{(a) SED of J004051. $\mathrm{\textit{T}_{bb}}$ is the temperature of the blackbody component introduced to fit the IR excess, and $\mathrm{\textit{T}_{star}}$ is the stellar effective temperature derived from the multi-band SED fitting. The black symbols mark the observed photometry, while the orange downward-pointing arrow indicates an upper limit. The red solid line is the blackbody component model required to fit the IR excess, and the blue solid line shows the combined model (8800 K stellar photosphere + 200 K blackbody). (b) Optical spectrum of J004051 (black) and the best-fit model (red). The spectral fitting results are shown in the title.}
\label{figure:J004051.58+403303.0_sed_specfit}
\end{figure}
% \vspace{-12mm} % 在这里插入10mm的垂直空间
% \vspace{-10mm} % 在这里插入10mm的垂直空间
\begin{figure}[htbp]
\centering

\subfloat[]{\includegraphics[scale=0.4]{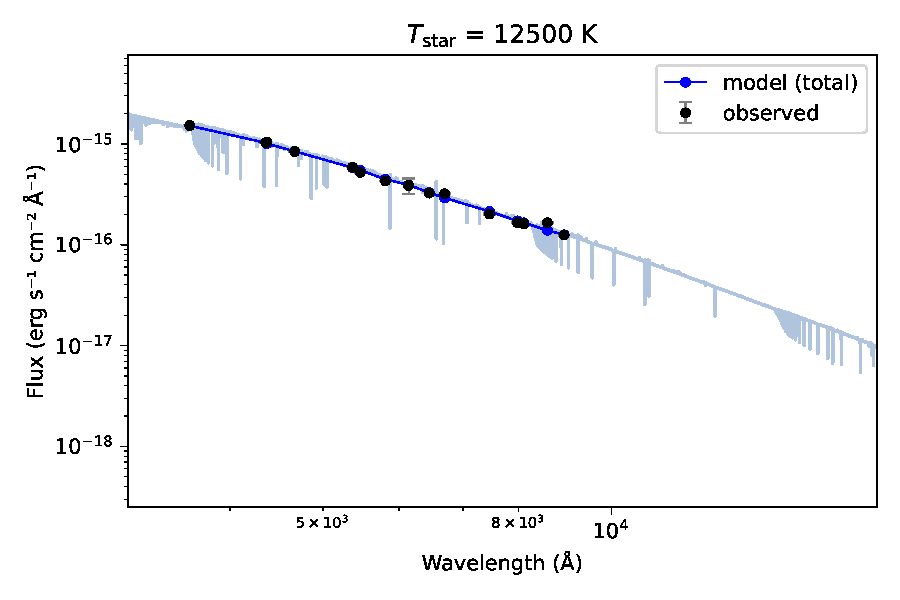}}
% \hfill
\subfloat[]{\includegraphics[scale=0.55]{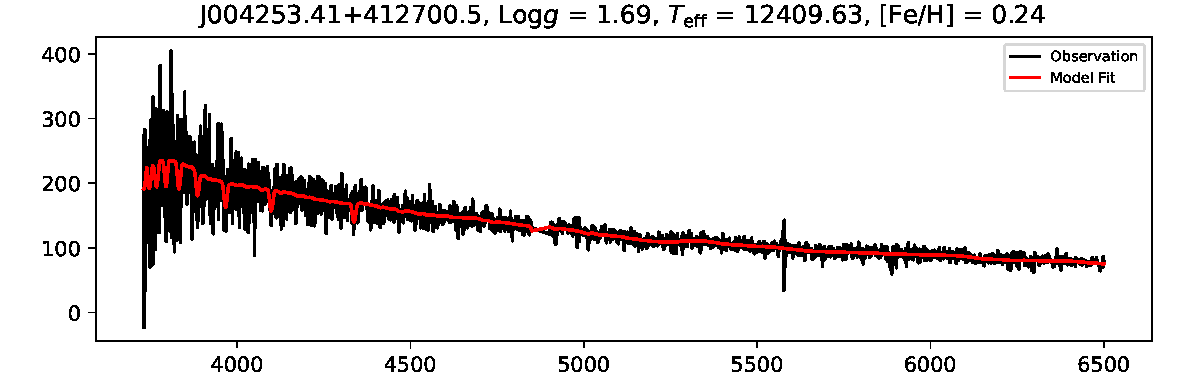}}

% \subfloat[]{\includegraphics[scale=0.4]{J004253_SED.pdf}}
% % \hfill
% \subfloat[]{\includegraphics[scale=0.55]{J004253.41+412700.5_specfit.pdf}}
\caption{The same with Fig.\,C1}
\label{figure:J004253.41+412700.5_sed_specfit}
\end{figure}
% \vspace{-12mm} % 在这里插入10mm的垂直空间
% \vspace{-10mm} % 在这里插入10mm的垂直空间
\begin{figure}[htbp]
\centering

\subfloat[]{\includegraphics[scale=0.4]{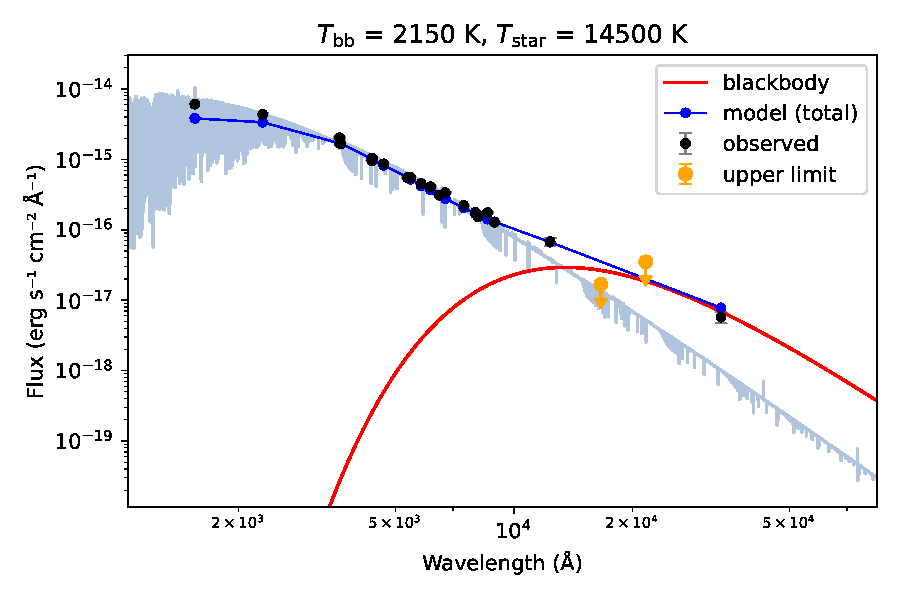}}
% \hfill
\subfloat[]{\includegraphics[scale=0.55]{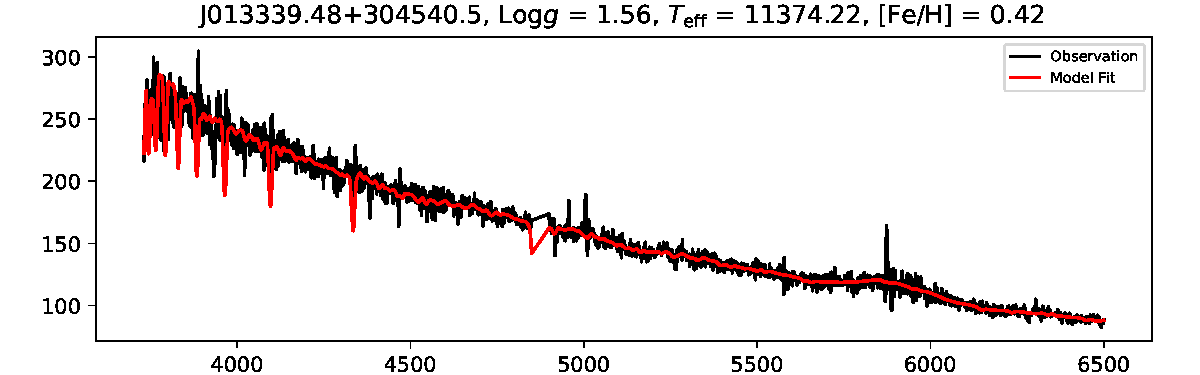}}

% \subfloat[]{\includegraphics[scale=0.4]{J013339_SED.pdf}}
% % \hfill
% \subfloat[]{\includegraphics[scale=0.55]{J013339.48+304540.5_specfit.pdf}}
\caption{The same with Fig.\,C1. But note that the binary nature of this source can account for the apparent temperature discrepancy between the SED-based and spectroscopic fitting results.}
\label{figure:J013339.48+304540.52_sed_specfit}
\end{figure}

% \vspace{-30mm} % 在这里插入10mm的垂直空间
% \vspace{-12mm} % 在这里插入10mm的垂直空间
\begin{figure}[htbp]
\centering

\subfloat[]{\includegraphics[scale=0.4]{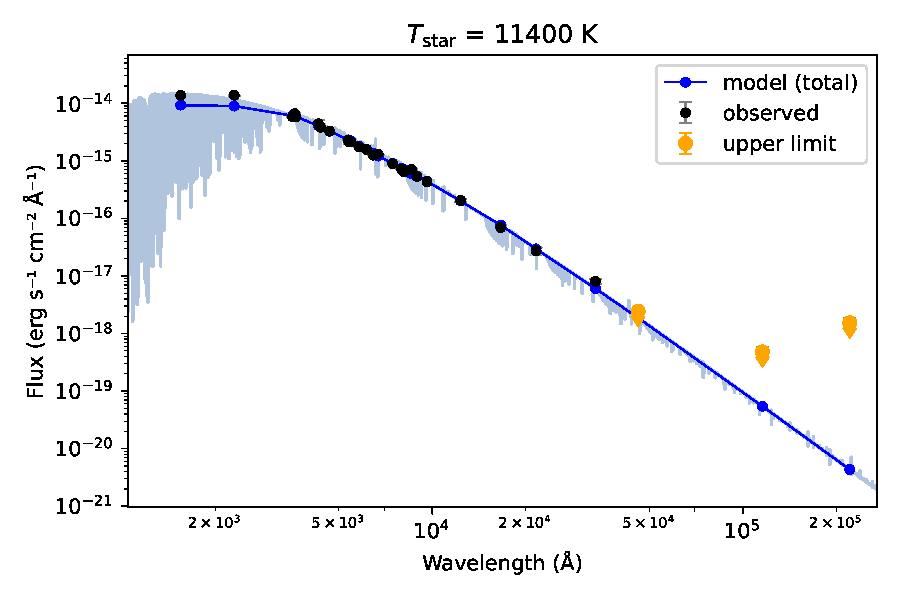}}
% \hfill
\subfloat[]{\includegraphics[scale=0.55]{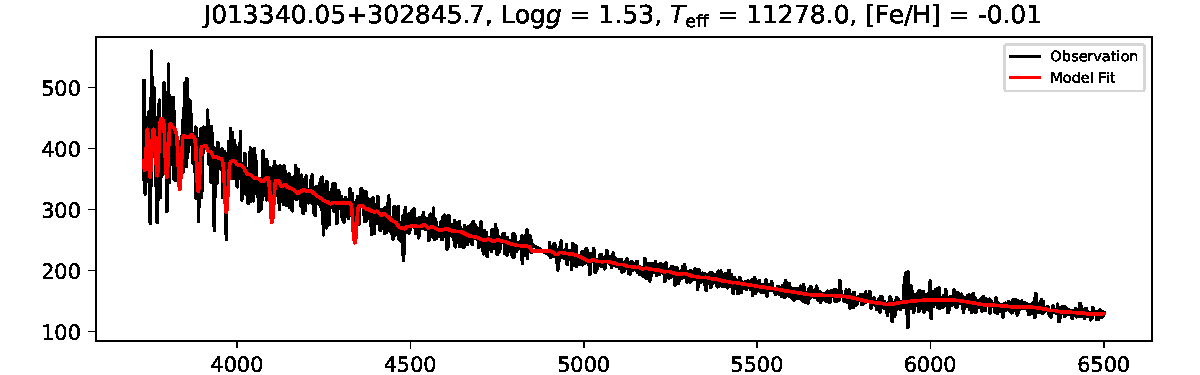}}

% \subfloat[]{\includegraphics[scale=0.4]{J013340_SED.pdf}}
% % \hfill
% \subfloat[]{\includegraphics[scale=0.55]{J013340.05+302845.7_specfit.pdf}}
\caption{The same with Fig.\,C1.}
\label{figure:J013340.05+302845.7_sed_specfit}
\end{figure}
% \vspace{-30mm} % 在这里插入10mm的垂直空间
% \vspace{-12mm} % 在这里插入10mm的垂直空间
\begin{figure}[htbp]
\centering
% \subfloat[]{\includegraphics[scale=0.4]{J013340_SED.pdf}}
% \hfill

\subfloat[]{\includegraphics[scale=0.55]{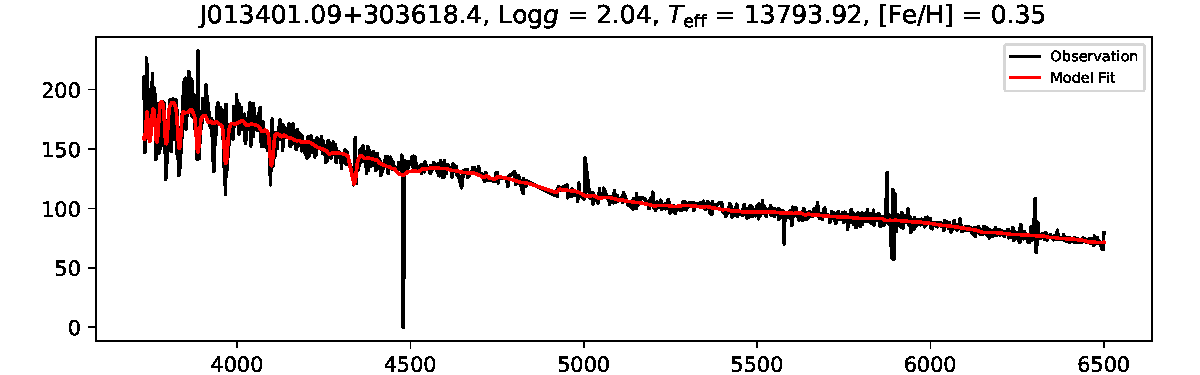}}

% \subfloat[]{\includegraphics[scale=0.55]{J013401.09+303618.4_specfit.pdf}}
\caption{The same with Fig.\,C1.}
\label{figure:J013401.09+303618.4_specfit}
\end{figure}

\begin{figure}[htbp]
\centering

\subfloat[]{\includegraphics[scale=0.4]{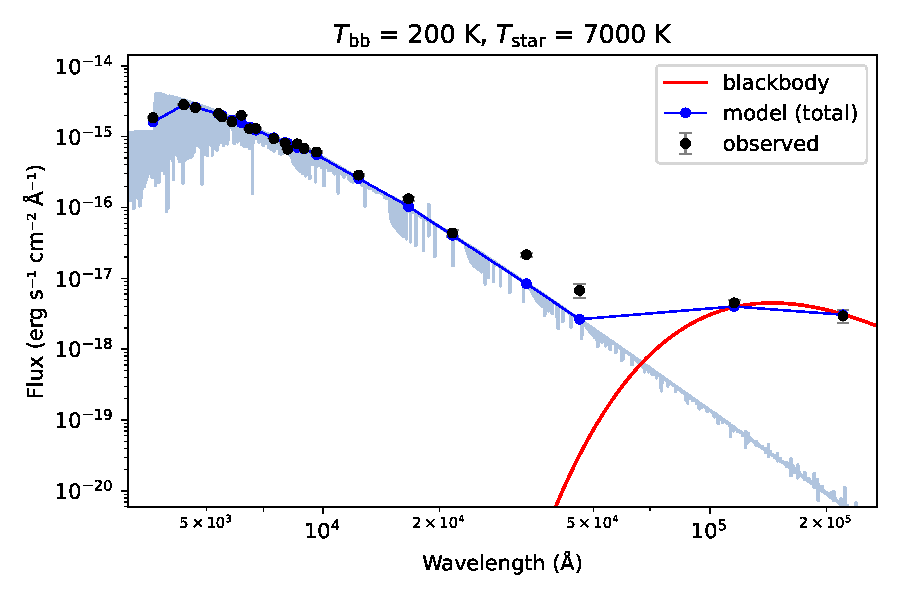}}
% \hfill
\subfloat[]{\includegraphics[scale=0.55]{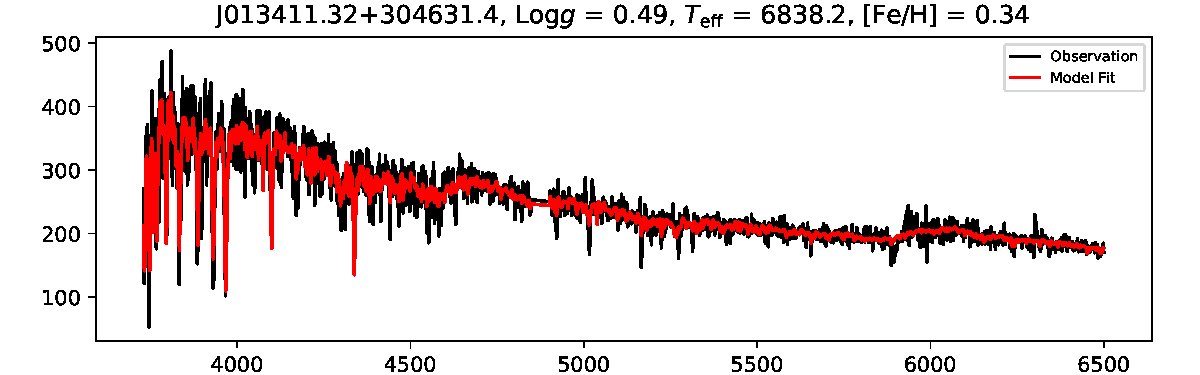}}

% \subfloat[]{\includegraphics[scale=0.4]{J013411_SED.pdf}}
% % \hfill
% \subfloat[]{\includegraphics[scale=0.55]{J013411.32+304631.4_specfit.pdf}}
\caption{The same with Fig.\,C1.}
\label{figure:J013411.32+304631.4_sed_specfit}
\end{figure}
% \vspace{-12mm} % 在这里插入10mm的垂直空间
% \vspace{-10mm} % 在这里插入10mm的垂直空间

\begin{figure}[htbp]
\centering

\subfloat[]{\includegraphics[scale=0.4]{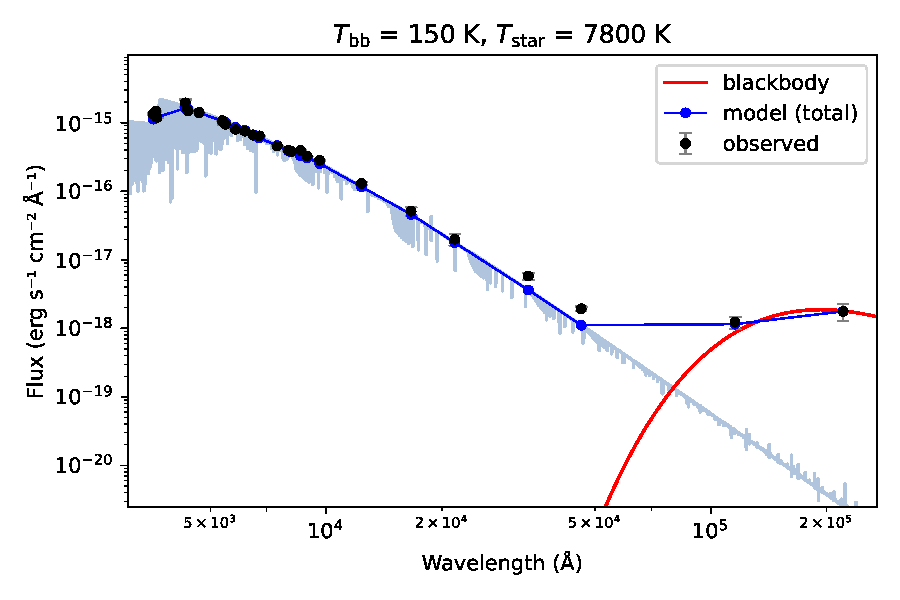}}
% \hfill
\subfloat[]{\includegraphics[scale=0.55]{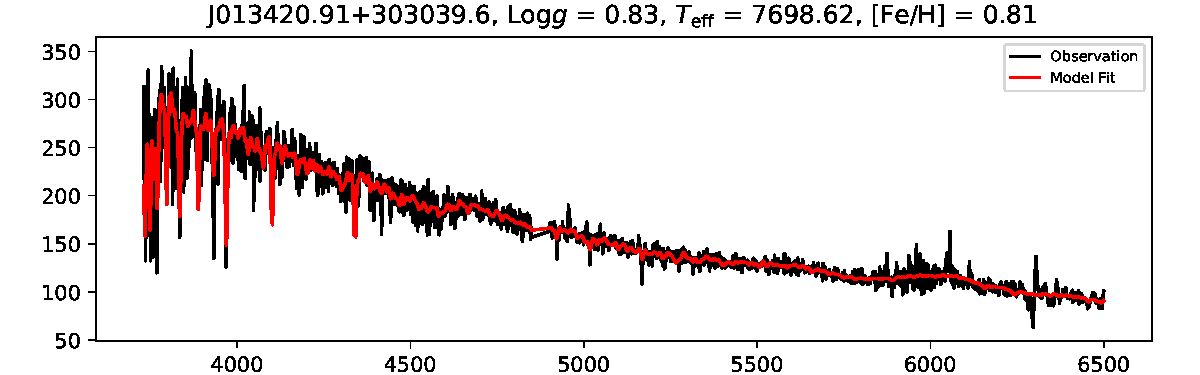}}

% \subfloat[]{\includegraphics[scale=0.4]{J013420_SED.pdf}}
% % \hfill
% \subfloat[]{\includegraphics[scale=0.55]{J013420.91+303039.6_specfit.pdf}}
\caption{The same with Fig.\,C1.}
\label{figure:J013420.91+303039.6_sed_specfit}
\end{figure}
% \vspace{-12mm} % 在这里插入10mm的垂直空间
% \vspace{-10mm} % 在这里插入10mm的垂直空间

\begin{figure}[htbp]
\centering
\subfloat[]{\includegraphics[scale=0.4]{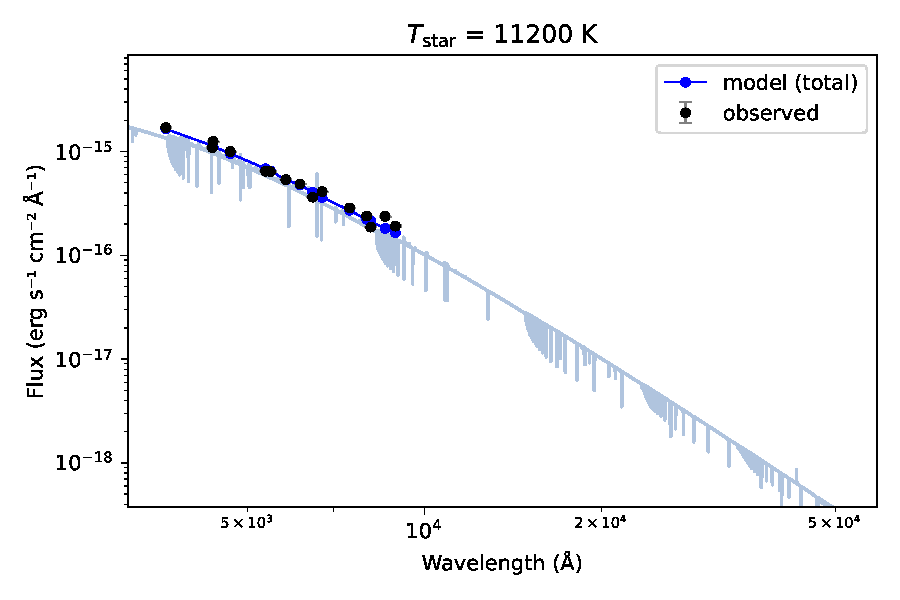}}
% \hfill
\subfloat[]{\includegraphics[scale=0.55]{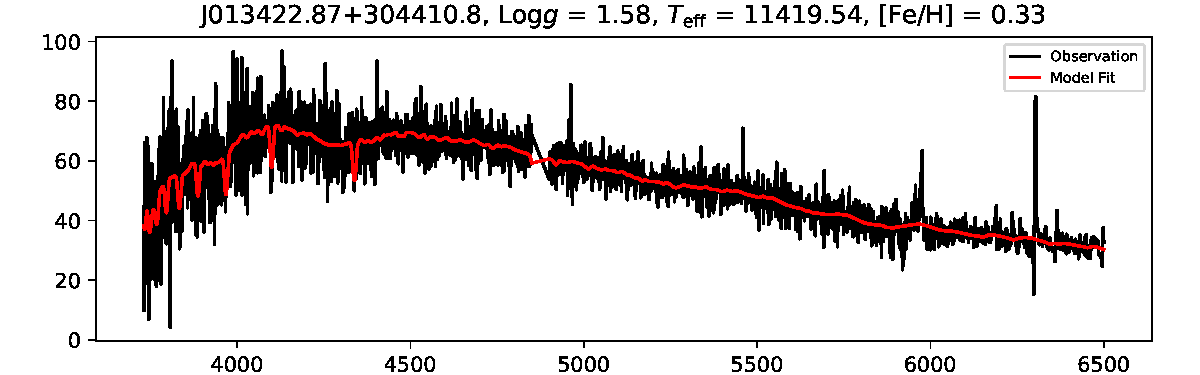}}

% \subfloat[]{\includegraphics[scale=0.4]{J013422_SED.pdf}}
% % \hfill
% \subfloat[]{\includegraphics[scale=0.55]{J013422.87+304410.8_specfit.pdf}}
\caption{The same with Fig.\,C1.}
\label{figure:J013422.87+304410.8_sed_specfit}
\end{figure}

% \label{lastpage}

\end{appendix}   
\end{CJK*}

\begin{thebibliography}{99}

\bibitem[Allard et al.(2013)]{2013MSAIS..24..128A} Allard, F., Homeier, D., Freytag, B., et al.\ 2013, MSAIS, 24, 128

\bibitem[Arneson et al.(2018)]{2018ApJ...864...31A} Arneson, R.~A., Shenoy, D., Smith, N., \& Gehrz, R.~D.\ 2018, \apj, 864, 1, 31

\bibitem[Bayo et al.(2008)]{2008A&A...492..277B} Bayo, A., Rodrigo, C., Barrado Y Navascu{\'e}s, D., et al.\ 2008, \aap, 492, 1, 277

\bibitem[Berkhuijsen et al.(1988)]{1988A&AS...76...65B} Berkhuijsen, E.~M., Humphreys, R.~M., Ghigo, F.~D., \& Zumach, W.\ 1988, \aaps, 76, 65

\bibitem[Chambers \& Pan-STARRS Team(2018)]{2018AAS...23110201C} Chambers, K., \& Pan-STARRS Team\ 2018, American Astronomical Society Meeting Abstracts, 231, 102.01

\bibitem[Clark et al.(2013)]{2013A&A...560A..10C} Clark, J.~S., Bartlett, E.~S., Coe, M.~J., et al.\ 2013, \aap, 560, A10

\bibitem[de Jager(1998)]{1998A&ARv...8..145D} de Jager, C.\ 1998, A\&A~Rev., 8, 1, 145

\bibitem[Deng et al.(2012)]{2012RAA....12..735D} Deng, L.-C., Newberg, H.~J., Liu, C., et al.\ 2012, \raa, 12, 6, 735

\bibitem[Drake et al.(2009)]{2009ApJ...696..870D} Drake, A.~J., Djorgovski, S.~G., Mahabal, A., et al.\ 2009, \apj, 696, 2, 870

\bibitem[Drout et al.(2012)]{2012ApJ...750...97D} Drout, M.~R., Massey, P., \& Meynet, G.\ 2012, \apj, 750, 1, 97

\bibitem[Ekstr{\"o}m et al.(2012)]{2012A&A...537A.146E} Ekstr{\"o}m, S., Georgy, C., Eggenberger, P., et al.\ 2012, \aap, 537, A146

\bibitem[Fabricant et al.(1998)]{1998SPIE.3355..285F} Fabricant, D.~G., Hertz, E.~N., Szentgyorgyi, A.~H., et al.\ 1998, SPIE Conf. Ser., 3355, 285

\bibitem[Fabricant et al.(2005)]{2005PASP..117.1411F} Fabricant, D., Fata, R., Roll, J., et al.\ 2005, \pasp, 117, 837, 1411

\bibitem[Flewelling et al.(2020)]{2020ApJS..251....7F} Flewelling, H.~A., Magnier, E.~A., Chambers, K.~C., et al.\ 2020, \apjs, 251, 1, 7

\bibitem[Fritz et al.(2012)]{2012A&A...546A..34F} Fritz, J., Gentile, G., Smith, M.~W.~L., et al.\ 2012, \aap, 546, A34

\bibitem[Gordon et al.(2016)]{2016ApJ...825...50G} Gordon, M.~S., Humphreys, R.~M., \& Jones, T.~J.\ 2016, \apj, 825, 1, 50

\bibitem[Graham et al.(2015)]{2015Natur.518...74G} Graham, M.~J., Djorgovski, S.~G., Stern, D., et al.\ 2015, \nat, 518, 7537, 74

\bibitem[Grassitelli et al.(2021)]{2021A&A...647A..99G} Grassitelli, L., Langer, N., Mackey, J., et al.\ 2021, \aap, 647, A99

\bibitem[Huang et al.(2019)]{2019ApJ...884L...7H} Huang, Y., Zhang, H.~W., Wang, C., et al.\ 2019, \apjl, 884, 1, L7

\bibitem[Humphreys \& Davidson(1994)]{1994PASP..106.1025H} Humphreys, R.~M., \& Davidson, K.\ 1994, \pasp, 106, 1025

\bibitem[Humphreys et al.(2013)]{2013ApJ...773...46H} Humphreys, R.~M., Davidson, K., Grammer, S., et al.\ 2013, \apj, 773, 1, 46

\bibitem[Humphreys et al.(2014)]{2014ApJ...790...48H} Humphreys, R.~M., Davidson, K., Gordon, M.~S., et al.\ 2014, \apj, 790, 48

\bibitem[Humphreys et al.(2015)]{2015PASP..127..347H} Humphreys, R.~M., Martin, J.~C., \& Gordon, M.~S.\ 2015, \pasp, 127, 950, 347

\bibitem[Humphreys et al.(2016)]{2016ApJ...825...64H} Humphreys, R.~M., Weis, K., Davidson, K., et al.\ 2016, \apj, 825, 1, 64

\bibitem[Humphreys et al.(2017a)]{2017ApJ...844...40H} Humphreys, R.~M., Davidson, K., Hahn, D., et al.\ 2017a, \apj, 844, 1, 40

\bibitem[Humphreys et al.(2017b)]{2017ApJ...836...64H} Humphreys, R.~M., Gordon, M.~S., Martin, J.~C., et al.\ 2017b, \apj, 836, 1, 64

\bibitem[Husser et al.(2013)]{2013A&A...553A...6H} Husser, T.~O., Wende-von Berg, S., Dreizler, S., et al.\ 2013, \aap, 553, A6

\bibitem[Ivanov et al.(1993)]{1993ApJS...89...85I} Ivanov, G.~R., Freedman, W.~L., \& Madore, B.~F.\ 1993, \apjs, 89, 85

\bibitem[Jayasinghe et al.(2019a)]{2019MNRAS.486.1907J} Jayasinghe, T., Stanek, K.~Z., Kochanek, C.~S., et al.\ 2019a, \mnras, 486, 2, 1907

\bibitem[Jayasinghe et al.(2019b)]{2019MNRAS.485..961J} Jayasinghe, T., Stanek, K.~Z., Kochanek, C.~S., et al.\ 2019b, \mnras, 485, 1, 961

\bibitem[Koposov et al.(2011)]{2011ApJ...736..146K} Koposov, S.~E., Gilmore, G., Walker, M.~G., et al.\ 2011, \apj, 736, 1, 146

\bibitem[Kostov \& Bonev(2018)]{2018BlgAJ..28....3K} Kostov, A., \& Bonev, T.\ 2018, Bulg. Astron. J., 28, 3

\bibitem[Kourniotis et al.(2017)]{2017A&A...601A..76K} Kourniotis, M., Bonanos, A.~Z., Yuan, W., et al.\ 2017, \aap, 601, A76

\bibitem[Kourniotis et al.(2018)]{2018MNRAS.480.3706K} Kourniotis, M., Kraus, M., Arias, M.~L., et al.\ 2018, \mnras, 480, 4, 3706

\bibitem[Kraus et al.(2014)]{2014ApJ...780L..10K} Kraus, M., Cidale, L.~S., Arias, M.~L., et al.\ 2014, \apj, 780, 1, L10

\bibitem[Liu et al.(2022)]{2022ApJ...932...29L} Liu, C., Kudritzki, R.-P., Zhao, G., et al.\ 2022, \apj, 932, 1, 29

\bibitem[Liu et al.(2014)]{2014IAUS..298..310L} Liu, X.~W., Yuan, H.~B., Huo, Z.~Y., et al.\ 2014, IAU Symp., 298, 310

\bibitem[Luo et al.(2012)]{2012RAA....12.1243L} Luo, A.~L., Zhang, H.-T., Zhao, Y.-H., et al.\ 2012, \raa, 12, 9, 1243

\bibitem[Maeder \& Meynet(2000)]{2000ARA&A..38..143M} Maeder, A., \& Meynet, G.\ 2000, \araa, 38, 1, 143

\bibitem[Magnier et al.(1992)]{1992A&AS...96..379M} Magnier, E.~A., Lewin, W.~H.~G., van Paradijs, J., et al.\ 1992, \aaps, 96, 379

\bibitem[Martin et al.(2023)]{2023RNAAS...7...96M} Martin, J.~C., Humphreys, R.~M., Weis, K., \& Bomans, D.~J.\ 2023, RNAAS, 7, 5, 96

\bibitem[Massey et al.(2006)]{2006AJ....131.2478M} Massey, P., Olsen, K.~A.~G., Hodge, P.~W., et al.\ 2006, \aj, 131, 5, 2478

\bibitem[Massey et al.(2007)]{2007AJ....134.2474M} Massey, P., McNeill, R.~T., Olsen, K.~A.~G., et al.\ 2007, \aj, 134, 6, 2474

\bibitem[Massey et al.(2016)]{2016AJ....152...62M} Massey, P., Neugent, K.~F., \& Smart, B.~M.\ 2016, \aj, 152, 3, 62

\bibitem[Neugent et al.(2012)]{2012ApJ...759...11N} Neugent, K.~F., Massey, P., \& Georgy, C.\ 2012, \apj, 759, 1, 11

\bibitem[Oksala et al.(2013)]{2013A&A...558A..17O} Oksala, M.~E., Kraus, M., Cidale, L.~S., et al.\ 2013, \aap, 558, A17

\bibitem[Richardson \& Mehner(2018)]{2018RNAAS...2..121R} Richardson, N.~D., \& Mehner, A.\ 2018, RNAAS, 2, 3, 121

\bibitem[Sarkisyan et al.(2020)]{2020MNRAS.497..687S} Sarkisyan, A., Sholukhova, O., Fabrika, S., et al.\ 2020, \mnras, 497, 1, 687

\bibitem[Sarkisyan et al.(2022)]{2022RAA....22a5022S} Sarkisyan, A., Sholukhova, O., Fabrika, S., et al.\ 2022, \raa, 22, 1, 015022

\bibitem[Sholukhova et al.(2015)]{2015MNRAS.447.2459S} Sholukhova, O., Bizyaev, D., Fabrika, S., et al.\ 2015, \mnras, 447, 3, 2459

\bibitem[Skrutskie et al.(2006)]{2006AJ....131.1163S} Skrutskie, M.~F., Cutri, R.~M., Stiening, R., et al.\ 2006, \aj, 131, 2, 1163

\bibitem[Smith(2017)]{2017RSPTA.37560268S} Smith, N.\ 2017, RSPTA, 375, 2105, 20160268

\bibitem[Solovyeva et al.(2019)]{2019MNRAS.484L..24S} Solovyeva, Y., Vinokurov, A., Fabrika, S., et al.\ 2019, \mnras, 484, 1, L24

\bibitem[Solovyeva et al.(2020)]{2020MNRAS.497.4834S} Solovyeva, Y., Vinokurov, A., Sarkisyan, A., et al.\ 2020, \mnras, 497, 4, 4834

\bibitem[Solovyeva et al.(2023)]{2023MNRAS.518.4345S} Solovyeva, Y., Vinokurov, A., Tikhonov, N., et al.\ 2023, \mnras, 518, 4, 4345

\bibitem[Ustamujic et al.(2021)]{2021A&A...654A.167U} Ustamujic, S., Orlando, S., Miceli, M., et al.\ 2021, \aap, 654, A167

\bibitem[van Genderen(2001)]{2001A&A...366..508V} van Genderen, A.~M.\ 2001, \aap, 366, 2, 508

\bibitem[Wenger et al.(2000)]{2000A&AS..143....9W} Wenger, M., Ochsenbein, F., Egret, D., et al.\ 2000, \aaps, 143, 1, 9

\bibitem[Wofford et al.(2020)]{2020MNRAS.493.2410W} Wofford, A., Ram{\'i}rez, V., Lee, J.~C., et al.\ 2020, \mnras, 493, 2, 2410

\bibitem[Wolf \& Stahl(1990)]{1990A&A...235..340W} Wolf, B., \& Stahl, O.\ 1990, \aap, 235, 1, 340

\bibitem[Zickgraf et al.(1996)]{1996A&A...315..510Z} Zickgraf, F.~J., Humphreys, R.~M., Lamers, H.~J.~G.~L.~M., et al.\ 1996, \aap, 315, 510

\end{thebibliography}
\end{document}